\documentclass{aa}
\usepackage{graphicx}
\usepackage{txfonts}
\usepackage{Multicols}
\begin{document}
   \title{Quantitative spectroscopy of Deneb  \thanks{Based on observations
   collected at the Centro Astron\'omico His\-pano Alem\'an (CAHA) at Calar Alto, 
   operated jointly by the Max-Planck Institut f\"ur Astronomie and the Instituto 
   de Astrofisica de Andalucia (CSIC).}}

   \author{F. Schiller
          \and
          N. Przybilla
          }

   \offprints{F.~Schiller, schiller@sternwarte.uni-erlangen.de}

   \institute{Dr. Remeis-Sternwarte Bamberg,
              Sternwartstr. 7, D-96049 Bamberg, Germany \\}

  \date{Received ... ; accepted ...}

 
  \abstract
{Quantitative spectroscopy of luminous BA-type supergiants offers a high potential for 
modern astrophysics. Detailed studies allow the evolution of massive stars and galactochemical evolution and permits the cosmic distance
scale to be constrained observationally.}
{A detailed and comprehensive understanding of the atmospheres of BA-type supergiants
is required in order to use this potential properly. The degree to which 
we can rely on quantitative
studies of this class of stars as a whole depends on the quality of the analyses for 
benchmark objects. We constrain the basic atmospheric parameters and fundamental stellar parameters, as well as chemical abundances of the prototype A-type supergiant Deneb to unprecedented accuracy by applying a sophisticated analysis methodology, which has recently been developed and tested.}
{The analysis is based on high-S/N and high-resolution spectra in the visual and near-IR.
Stellar parameters and abundances for numerous astrophysically
interesting elements are derived from synthesis of the photospheric spectrum 
using a hybrid non-LTE technique, i.e. line-blanketed LTE model atmospheres
and non-LTE line formation. Multiple metal ionisation equilibria and numerous hydrogen lines
from the Balmer, Paschen, Brackett, and Pfund series are made to match
simultaneously for the stellar parameter determination. The stellar wind
properties are derived from H$\alpha$ line-profile fitting using
line-blanketed hydrodynamic non-LTE models. Further constraints come from matching 
the photospheric spectral energy distribution from the UV to the near-IR $L$~band.}
{The atmospheric parameters of Deneb are tightly constrained: 
effective temperature $T_{\rm eff}$\,$=$\,8525\,$\pm$\,75\,K, surface gravity
$\log g$\,$=$\,1.10\,$\pm$\,0.05, microturbulence $\xi$\,$=$\,8\,$\pm$\,1\,km\,s$^{-1}$,
macroturbulence, and projected rotational velocity $v\,\sin\,i$ are both 
20\,$\pm$\,2\,km\,s$^{-1}$. The abundance analysis gives helium enrichment
by 0.10\,dex relative to solar and an N/C ratio of 4.44\,$\pm$\,0.84 (mass
fraction), implying strong mixing with CN-processed matter.
The heavier elements are consistently underabundant by $\sim$0.20\,dex compared to solar. Peculiar abundance patterns, which were derived in previous analyses to exist in Deneb, cannot be confirmed. Accounting for non-LTE effects is essential for removing systematic trends in
the abundance determination, for minimising statistical 1$\sigma$-uncertainties 
to $\lesssim$10-20\% and for establishing all ionisation equilibria at the same
time.}
{A luminosity of (1.96\,$\pm$\,0.32)$\times$10$^5$\,L$_{\odot}$, a radius of 
203\,$\pm$\,17\,R$_\odot$, and a current mass of 19\,$\pm$\,4\,M$_{\odot}$ are
derived. Comparison with stellar evolution predictions suggests that Deneb 
started as a fast-rotating late O-type star with $M^{\rm ZAMS}\simeq 23$\,M$_\odot$ 
on the main sequence and is currently evolving to the red supergiant stage.}


   \maketitle
%

\section{Introduction}
\label{introduction}

Massive stars are fundamental for the energy and momentum 
balance of galaxies. They represent important sources of ionising radiation and 
mass outflow through stellar winds. The final supernova explosions
contribute to the dynamics of the interstellar medium (ISM), stimulate star formation,
and enrich the ISM with heavy elements. Massive stars are therefore
major drivers of the cosmic cycle of matter. Moreover, because of their high
luminosities they can be observed over long distances, consequently
providing powerful indicators for the studying stellar and galactochemical
evolution in a wide variety of galactic environments.

Supergiants of spectral types B and A (BA-type supergiants) are among the 
visually brightest massive stars. Therefore, they are particularly interesting for
extragalactic astronomy. Spectroscopy of individual BA-type supergiants is
feasible in galaxies well beyond the Local Group using large ground-based
telescopes (Bresolin et al.~\cite{bresolin2001},~\cite{bresolin2002}).
So far, stellar abundances were compared to those from \ion{H}{ii} regions, and abundance gradients were measured (Urbaneja et al.~\cite{urbaneja2005}).
Moreover, BA-type supergiants show a high potential as standard candles 
for distance determinations, via application of the flux-weighted 
gravity-luminosity relationship (FGLR, Kudritzki et al.~\cite{kudritzki2003}) or 
via the wind momentum-luminosity relationship (WLR, Kudritzki et al.~\cite{kudritzki1999}). 
A crucial advantage over classical indicators such as Cepheids
is the possibility for a direct determination of metallicity and 
interstellar reddening from the quantitative analysis, allowing the systematic error budget to be reduced.

A comprehensive understanding of the atmospheres of BA-type supergiants 
has to be developed as the basis for detailed quantitative analyses in order 
to use their high potential as versatile indicators of stellar and 
galactic evolution and as distance indicators. The degree to which we can rely on quantitative
analyses of BA-type supergiants in general depends on the quality and
self-consistency of analysing benchmark objects, stars with comprehensive
observational data of the highest quality. In this respect, the brightest
A-type supergiant \object{Deneb} (A2\,Ia) represents one of the crucial
test cases for model atmosphere analyses of this class of stars.

Deneb is the best-studied A-type supergiant. Nevertheless, 
quantitative analyses using model atmospheres, from the pioneering work of
Groth~(\cite{groth1961}) to the most recent study of
Aufdenberg et al.~(\cite{aufdenberg2002}), find no consensus on atmospheric
parameters and elemental abundances. Published values for the effective temperature 
range e.g. from 7635\,K (Blackwell et al.~\cite{blackwell1980}) to 10\,080\,K
(Burnashev \cite{burnashev1980}), for the surface gravity from $\log
g$\,$=$\,1 (Zverko \cite{zverko1971}) to 1.54 (Burnashev
\cite{burnashev1980}).

Our aim is to apply recently introduced analysis methodology for BA-type
supergiants (Przybilla~\cite{przybilla2002}; Przybilla et
al.~\cite{przybilla2006}) on visual and near-IR spectra of the benchmark star Deneb. 
Stellar parameters and elemental abundances may thus be derived with 
unprecedented accuracy. In comparison
with recent models of the evolution of massive stars accounting for
mass loss and rotation (e.g. Heger \&
Langer~\cite{heger2000}; Meynet \& Maeder~\cite{meynet2003}) and the presence
of magnetic fields (Heger et al.~\cite{heger2005}; Maeder \& Meynet~\cite{maeder2005}
), this will
allow the evolutionary status of Deneb to be discussed. In addition, 
the stellar wind properties of Deneb will be investigated quantitatively.

The paper is organised as follows. We describe the observational data in 
Sect.~\ref{observations} and the derivation of the
stellar parameters in Sect.~\ref{stellarparameters}, while the 
elemental abundance determination is presented in Sect.~\ref{abundances}.
Additional constraints from near-IR spectroscopy are considered in
Sect.~\ref{irspectroscopy}. We discuss the evolutionary status of Deneb in
Sect.~\ref{evolutionarystatus}. Finally, the main results are summarised 
in Sect.~\ref{summaryandconclusions}.

\begin{figure}
\centering
\vspace{-0.3cm}
	\includegraphics[width=0.48\textwidth]{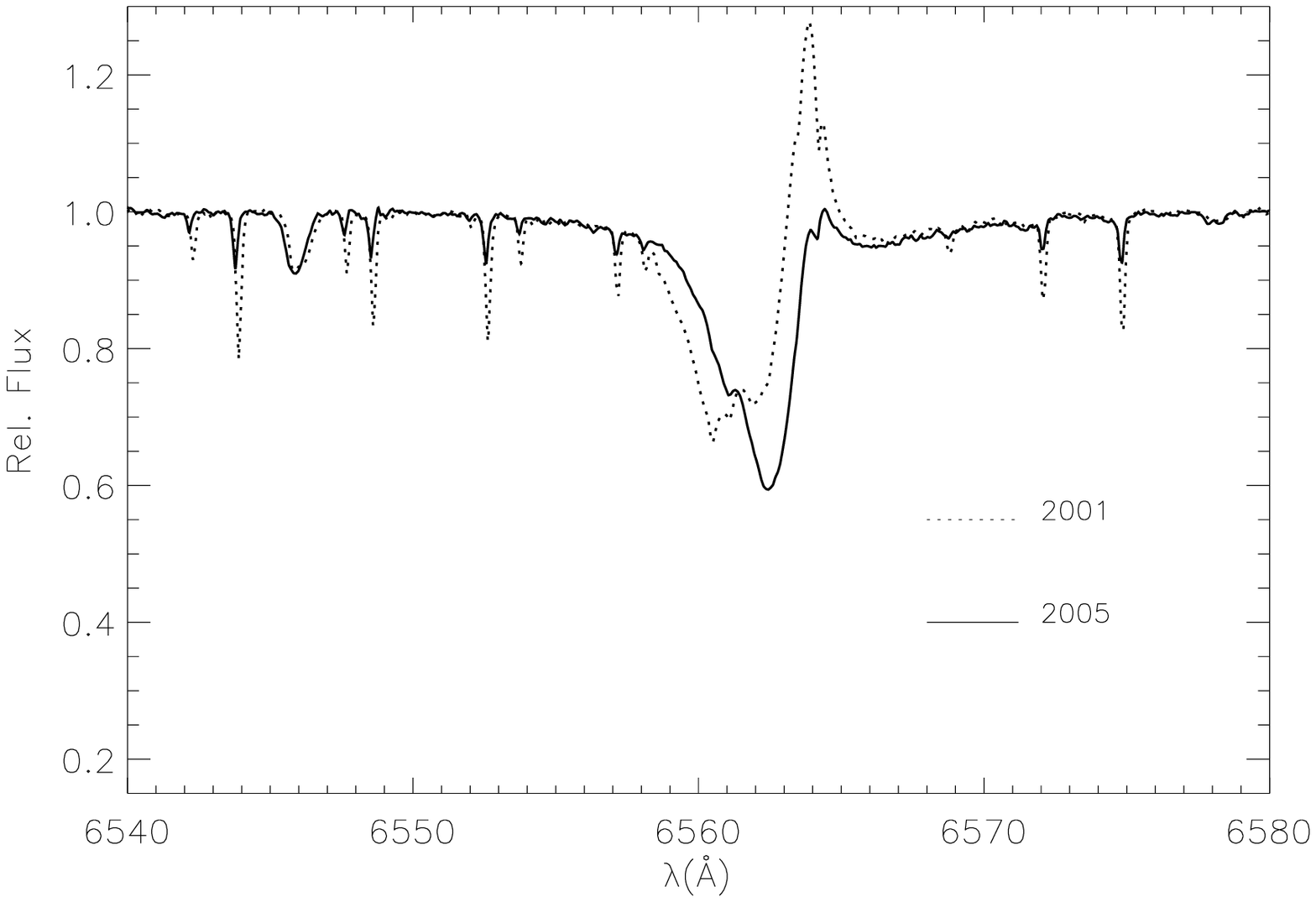}\\
\vspace{-0.3cm}
	\includegraphics[width=0.48\textwidth]{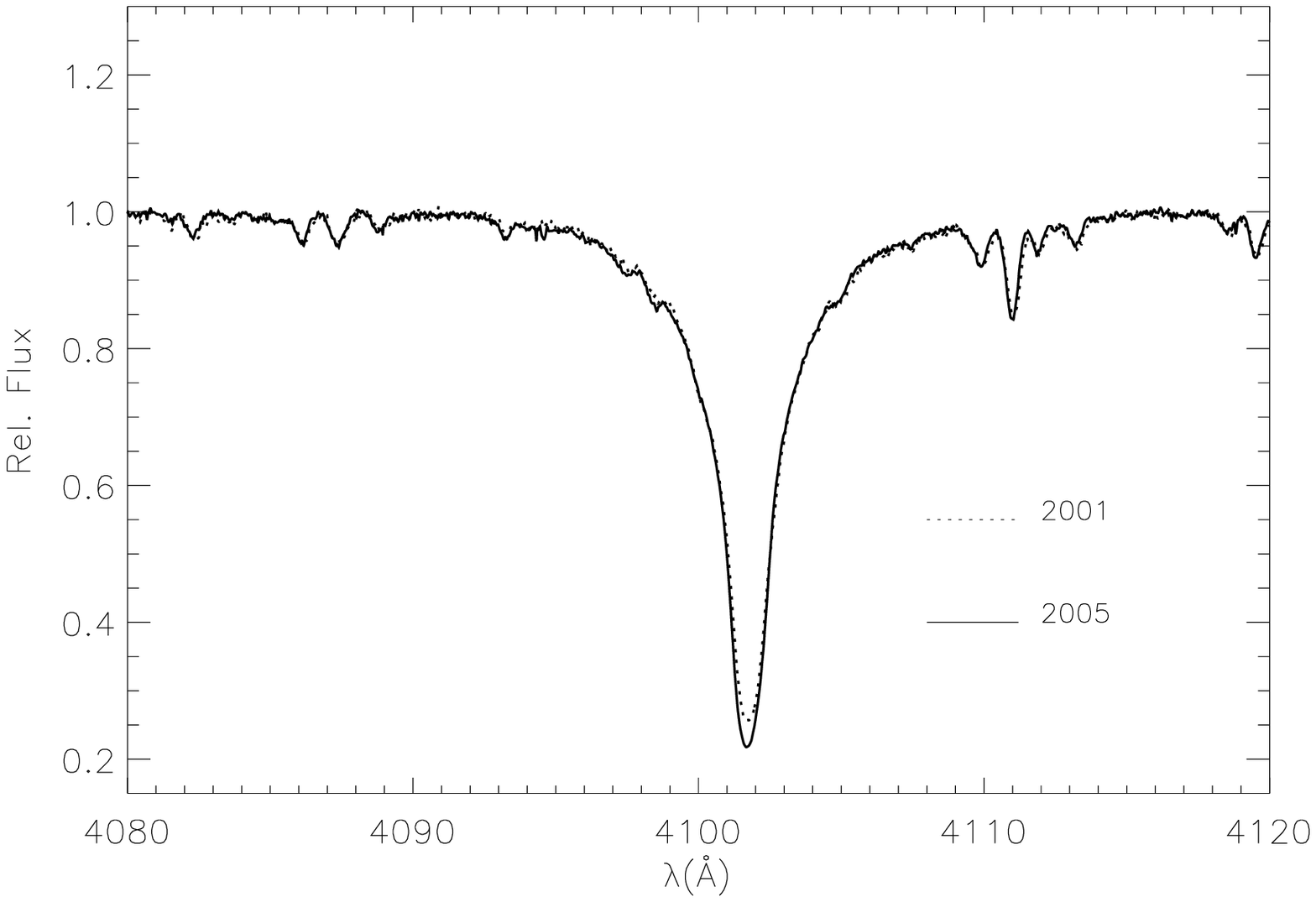}\\
\vspace{-0.3cm}
	\includegraphics[width=0.48\textwidth]{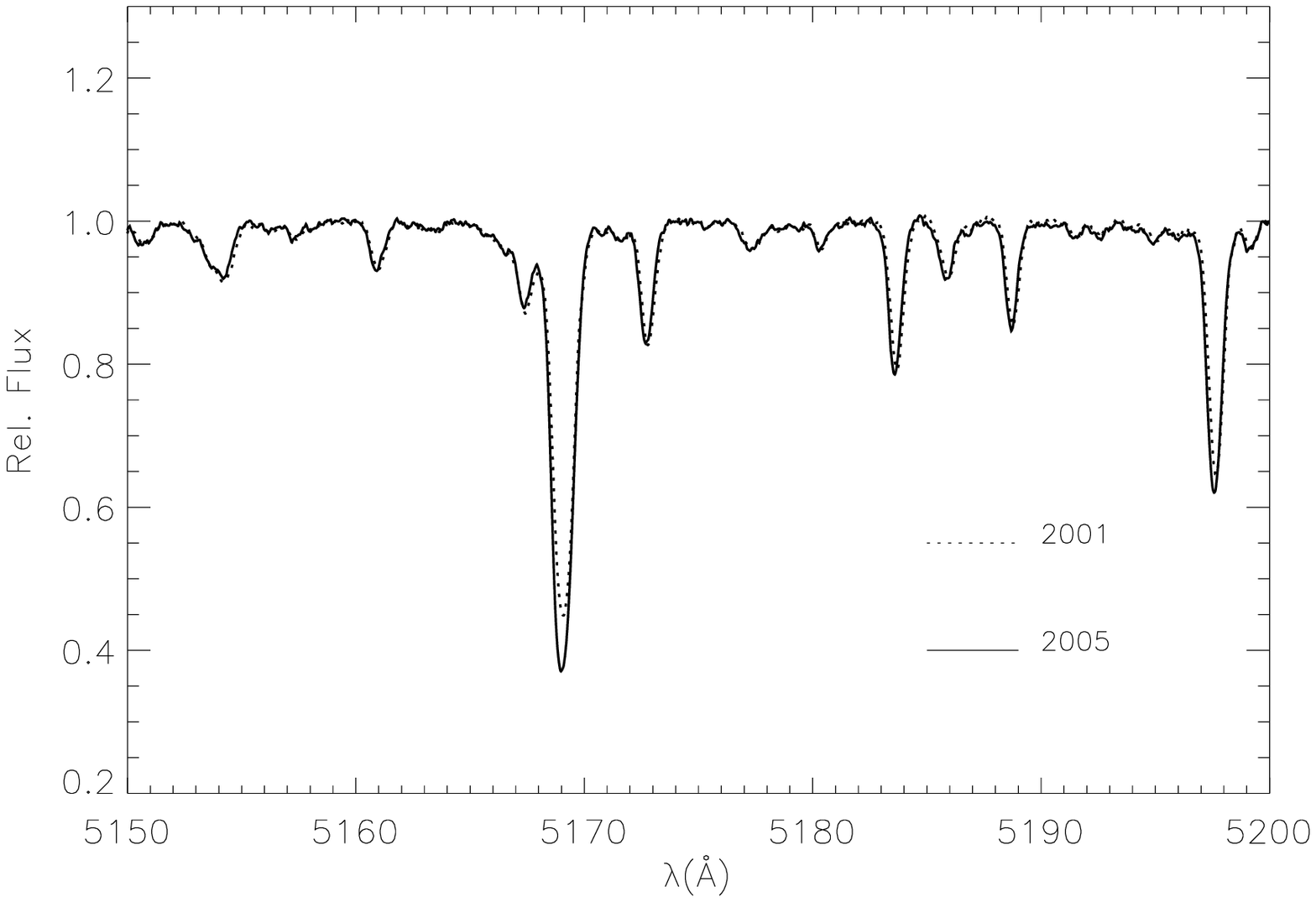}\\
\vspace{-0.3cm}
	\caption{Comparison between spectra of Deneb from 2001 (dotted) and 2005 (solid). Spectral regions around H$\alpha$ (\emph{upper}), H$\delta$ (\emph{middle}), and the \ion{Mg}{i} triplet (\emph{lower panel}) are displayed. Note the numerous sharp telluric H$_2$O features in the H$\alpha$ region. The wind-dominated H$\alpha$ line shows a pronounced P-Cygni profile in 2001 and reduced emission in 2005. Only the cores of the higher Balmer lines and few of the strongest metal lines (like \ion{Fe}{ii}\,$\lambda 5169$) are affected by the variable stellar wind.}
	\label{obscomparisons}
\end{figure}

\section{Observations and data reduction}
\label{observations}

Our primary observational data for the quantitative analysis are high-resolution,
high-S/N spectra taken on 21~September 2005 with the
Fibre Optics Cassegrain Echelle Spectrograph (FOCES,
Pfeiffer et al.~\cite{pfeiffer1998}) on the 2.2\,m-telescope at Calar Alto/Spain.
Eight spectra were obtained at 1\arcsec~seeing conditions
covering a wavelength range from 3860 to 9580\,{\AA} at resolving
power $R\,=\,\lambda /\Delta \lambda\,\approx$\,40\,000. After the reduction of each
spectrum using routines from the FOCES-package (correction for bad pixels and
cosmics, subtraction of bias and dark current, flatfielding, wavelength
calibration, rectification, and merging of all Echelle orders), the eight
spectra were coadded, resulting in a final S/N ratio of about 600
throughout most of the spectral range. Cross-correlation of this spectrum to
an appropriate synthetic spectrum at rest wavelength provides a correction 
for the radial velocity shift ($v_{\rm rad}=2.50$\,km\,s$^{-1}$). 
Comparison of our Echelle data (in particular around the Balmer lines) with a 
longslit spectrum\footnote{available via http://stellar.phys.appstate.edu/Standards/std1\_8.html} 
(at 1.8\,{\AA} resolution in the range from 3900 to 4600\,{\AA} taken at the Dark Sky Observatory, 
Appalachian State University) allows our continuum rectification to be
verified, giving excellent agreement. Also, a good agreement with the 
spectroscopic atlas of Albayrak et al.~{\cite{albayrak2003}} is found.

Besides the spectra of September 2005, we can consider data taken with the same instrument
on 26 September 2001. The comparison between these two spectra (Fig. \ref{obscomparisons}) shows 
a variable P-Cygni profile of H$\alpha$, while the higher Balmer
lines and the metal features remain almost unchanged. This implies that wind
variability will not affect the analysis of the photospheric features in
the present case. Our two spectra are representative of the states near maximum/minimum
wind emission reported by Kaufer et al.~(\cite{kaufer1996}) from a 2-year
long spectroscopic campaign on Deneb.

Photometric data, as well as UV and near-infrared 
spectra, are used to verify our analysis 
via spectral energy distribution fits and comparison with additional diagnostic
spectral features. Here, the data considered are
\begin{itemize}
\item [1.] flux-calibrated, low-dispersion spectra from 1150 to 1980\,{\AA} and
from 1850 to 3290\,{\AA}, obtained by the IUE satellite (SWP09133 and LWR07864, respectively,
as extracted from the INES archive\footnote{
http://ines.laeff.esa.es/}), 
\item [2.] $UBV$ photometry from Johnson et al.~(\cite{johnson1966}) and $RIL$ from 
Morel \& Magnenat (\cite{morel1978}), Str\"omgren photometry ($m_{1}$, $b-y$, $c_{1}$) 
from Hauck \& Mermilliod (\cite{hauck1998}), and $JHK$-data
from the Two Micron All Sky Survey (2MASS, Skrutskie et al.~\cite{skrutskie2006}),
\item [3.] high-quality spectra ($R \approx 3000-10\,000$, S/N\,$\sim$100) in the 
$Jw$-, $H$-, and $K$-band (Fullerton \& Najarro~\cite{fullerton1998}) and a $K$-band spectrum 
($R \approx 3000$) from Wallace \& Hinkle~(\cite{wallace1997}). 
\end{itemize}

\section{Stellar parameters}
\label{stellarparameters}

\subsection{Basic atmospheric parameters}
\label{atmosparameters}

The basic atmospheric parameters, such as effective temperature 
$T_{\rm eff}$, surface gravity $\log\,g$, microturbulent velocity $\xi$, 
helium abundance $n$(He), together with macroturbulent velocity $\zeta$ and 
projected rotational velocity $v \sin i$, are derived from a comparison
of observation with model spectra. The synthetic spectra are computed using both 
LTE and non-LTE techniques. 

The LTE model atmospheres are calculated with the ATLAS9 code (Kurucz~\cite{kurucz1993}), 
in the version of M. Lemke as obtained from the CCP7 software library. Further 
modifications (Przybilla et al.~\cite{przybilla2001b}) allow model convergence
close to the Eddington limit to be obtained, which turns out to be crucial
in the case of Deneb. The LTE model atmospheres are the basis for non-LTE 
line-formation computations with DETAIL and SURFACE 
(Giddings~\cite{giddings1981}; Butler \& Giddings~\cite{butler1985};
recently updated by K.~Butler). 
Non-LTE model atoms according to Table~\ref{nltemodelatoms} are considered. 
This hybrid non-LTE technique is able to reproduce observation at high
quality with modest requirements for computing time, see Przybilla et 
al.~(\cite{przybilla2006}) for a detailed discussion.
Line-formation calculations assuming LTE are also performed with SURFACE.

\begin{table}
\caption{Non-LTE model atoms.}
\label{nltemodelatoms}
\centering
\begin{tabular}{l l}
\hline\hline
Ion & Source \\
\hline
H & Przybilla \& Butler (\cite{przybilla2004})\\
\ion{He}{i} & Przybilla (\cite{przybilla2005}) \\
\ion{C}{i/ii} & Przybilla et al.~(\cite{przybilla2001b}), Nieva \& Przybilla (\cite{nieva2006}, \cite{nieva2007})\\
\ion{N}{i/ii} & Przybilla \& Butler (\cite{przybilla2001})\\
\ion{O}{i} & Przybilla et al.~(\cite{przybilla2000}) \\ 
\ion{Mg}{i/ii} & Przybilla et al.~(\cite{przybilla2001a})\\
\ion{S}{ii} & Vrancken et al.~(\cite{vrancken1996}), with updated atomic data\\
\ion{Ti}{ii} & Becker (\cite{becker1998})\\
\ion{Fe}{ii} & Becker (\cite{becker1998})\\
\hline
\end{tabular}
\end{table}

\begin{table}
\caption{Stellar parameters of Deneb.}
\label{allparameters}
\begin{center}
\begin{tabular}{lc}

\hline\hline

Name					& Deneb \\
Spectral type 				& A2\,Ia \\
$d$					& 802\,$\pm$\,66\,pc \\
Radial velocity				& 2.50\,km\,s$^{-1}$ \\[2mm]


\textbf{Atmosphere}	& \\

$T_{\rm eff}$		& 8525\,$\pm$\,75\,K \\
$\log g$ (cgs)		& 1.10\,$\pm$\,0.05 \\
$n$(He) (number fraction) & 0.11\,$\pm$\,0.01\\
$[M/H$]			& $-$0.20\,$\pm$\,0.04 \\ 
$\xi$			& 8\,$\pm$\,1\,km\,s$^{-1}$ \\
$\zeta $		& 20\,$\pm$\,2\,km\,s$^{-1}$\\
$v \sin{i}$		& 20\,$\pm$\,2\,km\,s$^{-1}$\\[2mm]


\textbf{Photometry}			& \\

$V ~^a$ 			& $1.^{\hspace{-2pt}m}25$ \\
$B - V ~^a$ 			& $+0.^{\hspace{-2pt}m}09$ \\
$U - B ~^a$ 			& $-0.^{\hspace{-2pt}m}23$ \\
$(B - V)_0 ~^b$			& $0.^{\hspace{-2pt}m}05$ \\
$E(B - V) $ 			& $0.^{\hspace{-2pt}m}04$ \\
$A_V $				& $0.^{\hspace{-2pt}m}11$ \\
$(m_V-M_V)_0 ~^c$		& $9.^{\hspace{-2pt}m}52 \pm 0.^{\hspace{-2pt}m}18$\\
$M_{V} $			& $-8.^{\hspace{-2pt}m}38 \pm 0.^{\hspace{-2pt}m}18$\\
$B.C. ~^b$			& $-0.^{\hspace{-2pt}m}11$ \\
$M_{\rm bol} $			& $-8.^{\hspace{-2pt}m}49 \pm 0.^{\hspace{-2pt}m}18$\\[2mm]

\textbf{Physical}			& \\

$\log\,L\,/\,L_\odot $			& 5.30\,$\pm$\,0.07\\
$R\,/\,R_\odot $			& 203\,$\pm$\,17\\
$M^{\rm spec}\,/\,M_\odot $		& 19\,$\pm$\,4\\
$M^{\rm ZAMS}\,/\,M_\odot $		& 23\,$\pm$\,2\\
$M^{\rm evol}\,/\,M_\odot $		& 18\,$\pm$\,2\\[2mm]

\textbf{Wind parameters} & \\

$\dot{M}$ & $3.1\cdot 10^{-7}\mbox{M}_\odot\,\mbox{yr}^{-1}$ \\
$v_\infty\,^d$ & $240$\,$\pm$25\,km\,s$^{-1}$ \\
$\beta$ & 3.0 \\
$\xi_{\mbox{H}\alpha}$ & $35\,\rm km\,s^{-1}$ \\[1mm]
\hline
\end{tabular}
\end{center}
$^a$ Johnson et al.~(\cite{johnson1966}); $^b$ calculated from the ATLAS9
model;\\ 
$^c$ Humphreys (\cite{humphreys1978}); $^d$ Hirsch~(\cite{hirsch1998})
\end{table}

Several spectral features can be utilised for determining the basic 
atmospheric parameters via fits of synthetic spectra to observation. 
Ionisation equilibria of metals like \ion{Mg}{i/ii}, \ion{C}{i/ii},
\ion{N}{i/ii} (and other species showing spectral lines from two or more
ionisation stages) are highly sensitive indicators for 
$T_{\rm eff}$, as well as for $\log g$ in early A-type supergiants. 
The hydrogen lines are mainly sensitive to variations in $\log g$. 
The helium abundance has to be determined simultaneously with the other 
atmospheric parameters in order to minimise systematic errors. An increase in 
helium abundance has similar effects on the model predictions as an 
increase in surface gravity, i.e. a pressure rise through the increase
in mean molecular weight (Kudritzki~\cite{kudritzki1973}). A list of spectral lines (helium and metals) 
analysed in the present study can be found in the Appendix\footnote{Appendix~A 
is only available in electronic form at http://www.edpsciences.org}. The 
photospheric hydrogen Balmer lines from H$_{\beta}$ to H$_8$
are analysed, and in addition numerous Paschen, Brackett, and Pfund 
lines. H$_\alpha$ is omitted in the derivation of the basic atmospheric parameters as it is strongly affected by wind emission.

\begin{figure*}
\centering
	\includegraphics[width=0.48\textwidth]{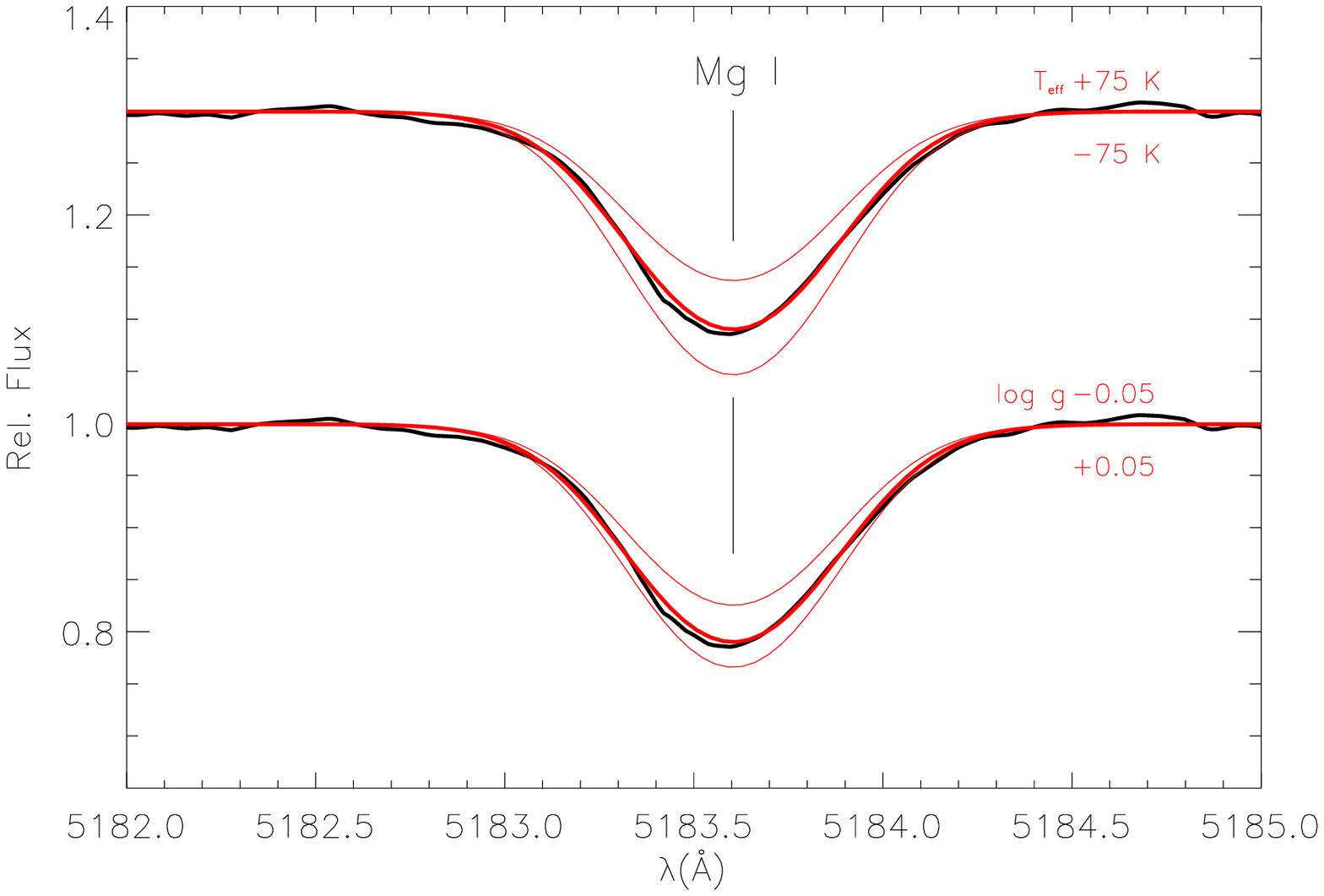}
	\includegraphics[width=0.48\textwidth]{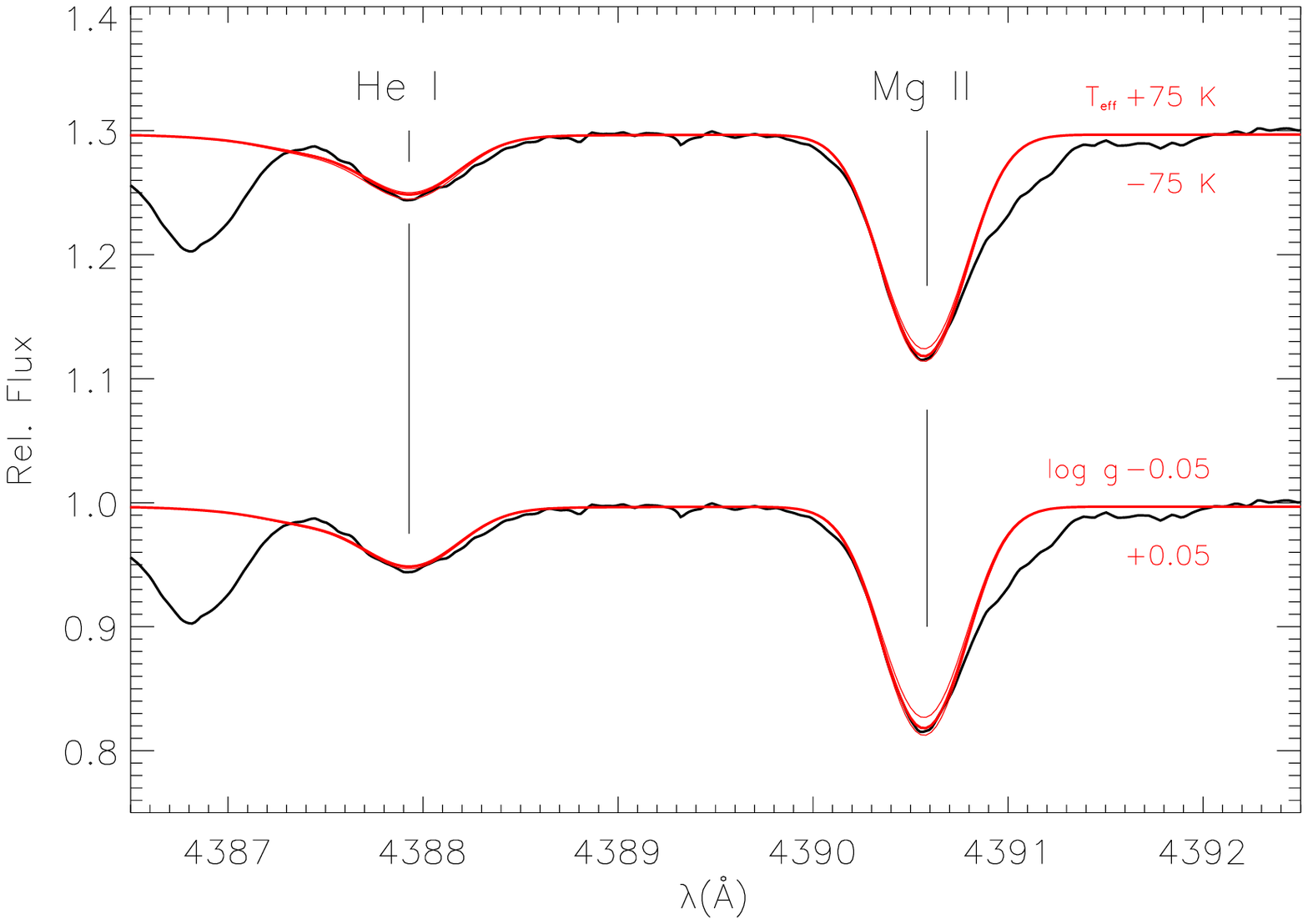}
\vspace{-0.4cm}
	\caption{Temperature determination for Deneb using the \ion{Mg}{i/ii} 
	non-LTE ionisation equilibrium. Displayed are the observed line profiles 
	for some strategic lines, \ion{Mg}{i}\,$\lambda$\,5183 (\emph{left panel}) and 
	\ion{Mg}{ii}\,$\lambda$\,4390 and \ion{He}{i}\,$\lambda$\,4387
	(\emph{right panel}), and the best fit for stellar parameters as given in Table \ref{allparameters} (thick red line); theoretical profiles for varied parameters are also shown (thin red lines, as indicated). A vertical shift by 0.3 units has been applied to the upper set of profiles. Note the strong sensitivity of the minor ionic species, \ion{Mg}{i}, to parameter changes, while \ion{Mg}{ii} is virtually unaffected.}
	\label{hemg}
\end{figure*}

\begin{figure*}
\centering
	\includegraphics[width=0.48\textwidth]{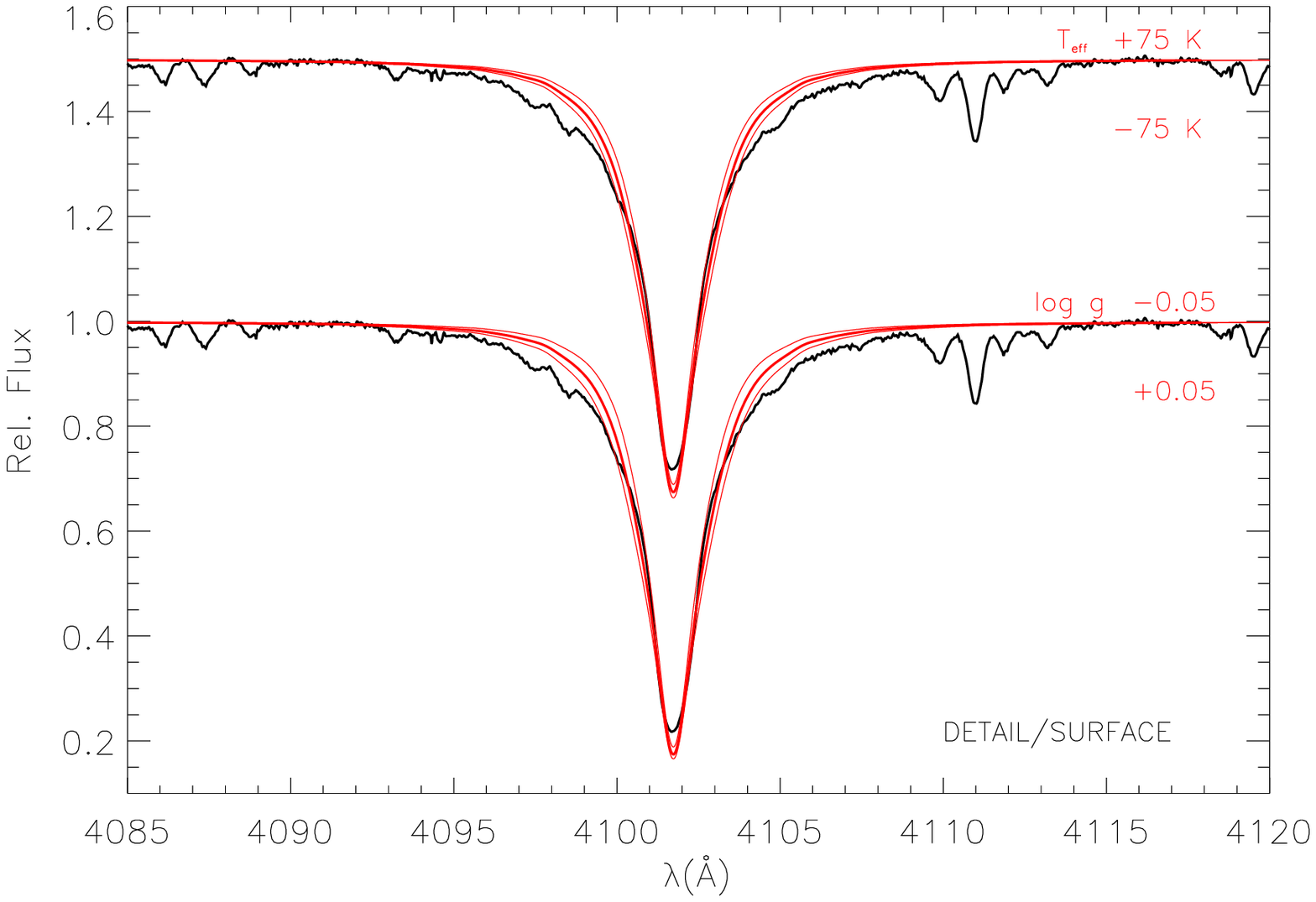}
	\includegraphics[width=0.48\textwidth]{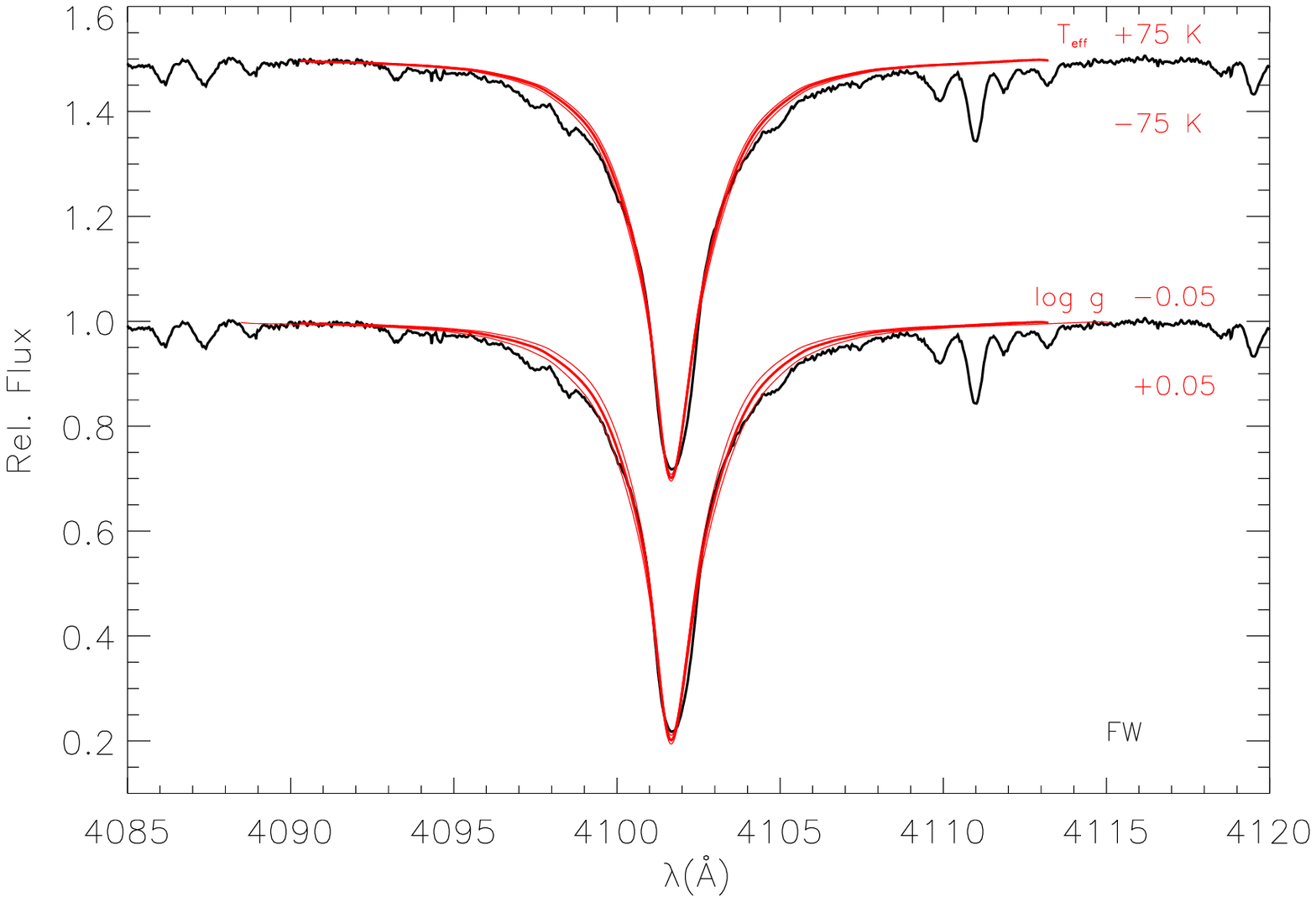}
\vspace{-0.4cm}
	\caption{Impact of stellar parameter variations on non-LTE line
	profile fits (\emph{left panel}). Synthetic spectra for the adopted
	parameters (see Table~\ref{allparameters}, thick red line) and varied parameters as indicated (thin red lines) are compared to observations. A vertical shift by 0.5 units has been applied to the upper profiles. \emph{Right panel}: observation and spectrum synthesis with FASTWIND for $\log g$\,=\,1.12\,dex (central line) and two variations as indicated. Note that, although the hydrodynamical FASTWIND model produces a better agreement with the observation, the resulting parameters of both methods are practically identical.}
	\label{hdelta}
\end{figure*}

We derive the atmospheric parameters in an iterative process from our set of 
hydrogen, helium, magnesium, and nitrogen lines. The results (and further 
information on Deneb) are summarised in Table \ref{allparameters}. 
The agreement between theory and observation for these final parameters and the 
response of some of the diagnostic lines to variations at the amount 
of the derived uncertainties (75\,K in $T_{\rm eff}$ and 0.05 in $\log g$) 
are illustrated in Figs.~\ref{hemg} and~\ref{hdelta}.

\begin{table}
\caption{Stellar parameters of Deneb from different sources.}
\label{parametersliterature}
\centering
\begin{tabular}{l c c c}
\hline\hline
Source						& $T_{\rm eff}$ in K	& $\log g$	& $\xi$	in km\,s$^{-1}$	 \\
\hline
\emph{This work}				& 8525\,$\pm$\,75	& 1.10\,$\pm$0.05	& 8\,$\pm$\,1 \\[2mm]
Blackwell et al.~(\cite{blackwell1980})		& 7635			& --		& -- \\[1mm]
Bonneau et al.~					& 8250			& --		& -- \\
\hspace{0.5cm}	(\cite{bonneau1981})	       	& 8150\,$\pm$\,600	& --		& -- \\[1mm]
Aufdenberg et 					& 8600\,$\pm$\,500	& 1.3		& 15 \\
\hspace{0.5cm}	al. (\cite{aufdenberg2002})	& 8420\,$\pm$\,100	& 1.1$-$1.6	& -- \\[1mm]
Albayrak (\cite{albayrak2000})			& 9000			& 1.45		& 3.6$-$11.9 \\[1mm]
Takeda et al.~(\cite{takeda1996})			& 9000			& 1.5		& 10 \\[1mm]
Samedov (\cite{samedov1993})			& 9100			& 1.2		& -- \\[1mm]
Groth (\cite{groth1961})			& 9170\,$\pm$\,500	& 1.13\,$\pm$\,0.2	& -- \\[1mm]
Zverko (\cite{zverko1971})		& 9510			& 1		& 11 / 17 \\[1mm]
Aydin (\cite{aydin1972})			& 9900			& 1.2\,$\pm$\,0.2	& 5$-$12 \\[1mm]
Takeda (\cite{takeda1994})		& 10\,000		& 1.5		& 10 \\[1mm]
Burnashev (\cite{burnashev1980})		& 10\,080		& 1.54		& -- \\

\hline
\end{tabular}
\end{table}

The microturbulent velocity is determined primarily from the \ion{N}{i} 
lines on the condition that the abundance indicated by a line does not depend 
on its equivalent width. This condition is also fulfilled for all
elements with non-LTE calculations, as discussed later (Fig.~\ref{abuvsw}). 
Macroturbulence and $v \sin i$ are derived by detailed line-profile
fits to features of various elements.


\begin{figure*}
\centering
	\includegraphics[width=0.48\textwidth]{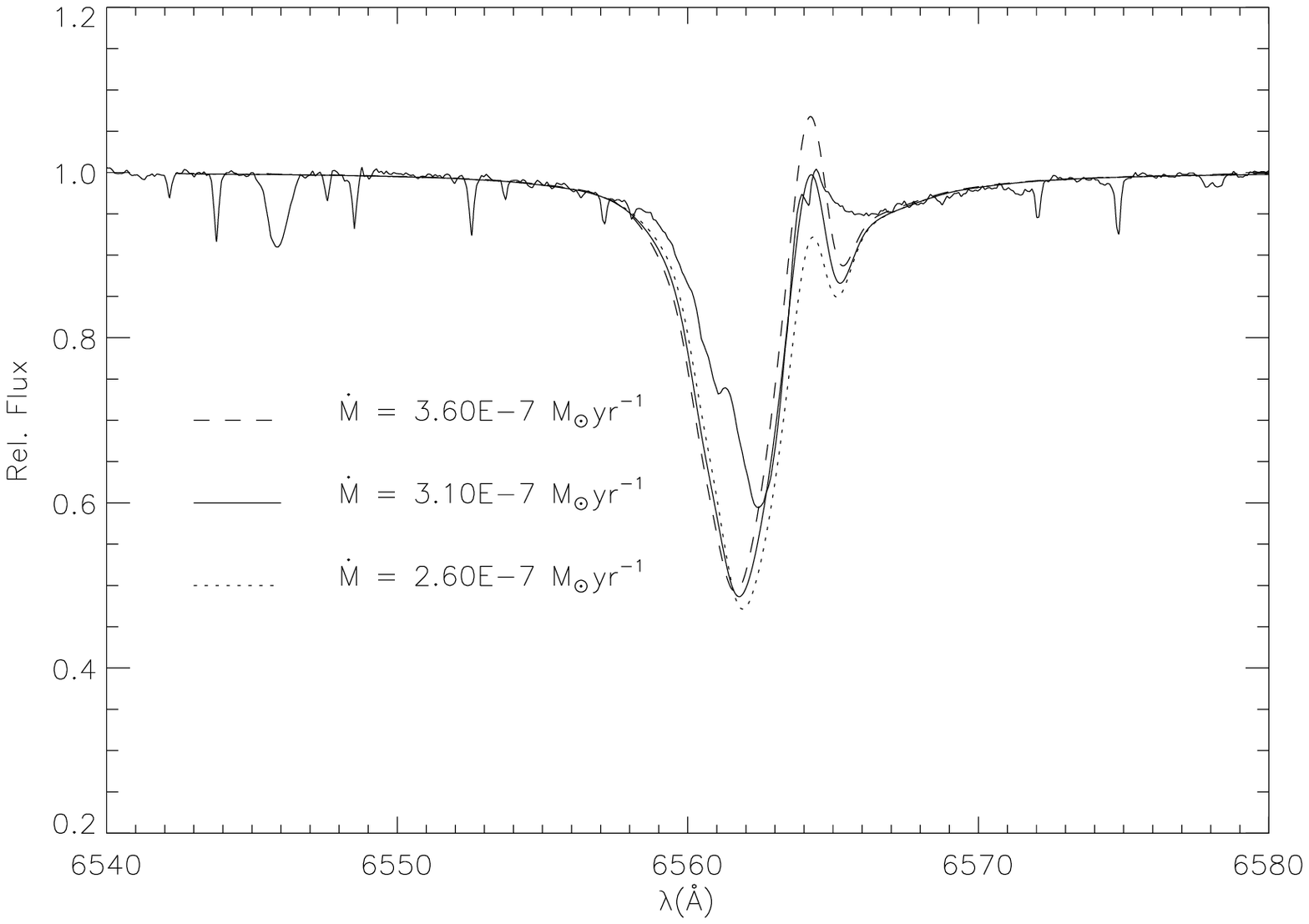}
	\includegraphics[width=0.48\textwidth]{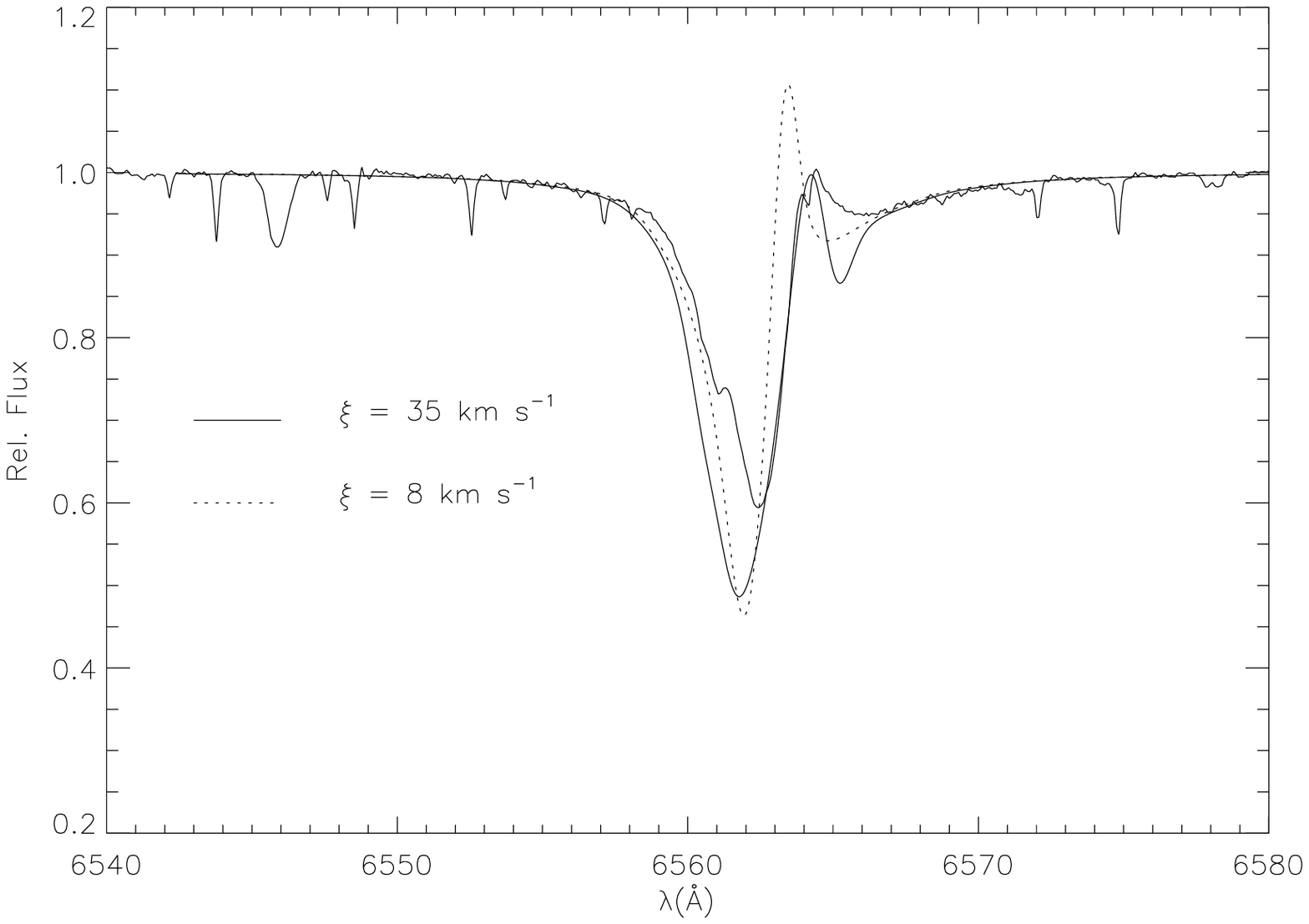}
\vspace{-0.3cm}
	\caption{Impact of variations in mass-loss rate and turbulent velocity on H$\alpha$ model spectra (computed with FASTWIND). The final model was calculated with the wind parameters in Table \ref{allparameters}\textbf{.}}
	\label{halphaFW}
\end{figure*}

Only a fair compromise could be achieved when fitting the cores and the wings of 
the Balmer lines simultaneously with our hybrid non-LTE technique. Figure~\ref{hdelta} 
(left panel) shows an example. This prompted us to perform test calculations 
with FASTWIND (Puls et al.~(\cite{puls2005}) in order to obtain
hydrodynamic, line-blanketed non-LTE models in spherical geometry. Specification of stellar mass-loss rate $\dot{M}$, wind end
velocity $v_\infty$, exponent of the velocity law $\beta$, and stellar radius
$R$ is required as input parameters. Improved matches
between models and observation are achieved, see Figs.~\ref{hdelta} 
and~\ref{infrared_overview}. The wind parameters can be constrained from H$\alpha$-profile
fitting, see e.g. McCarthy et al.~(\cite{mccarthy1997}) for details and
Fig.~\ref{halphaFW} for examples in the present case. The wind end
velocity is a mean value from the analysis of UV resonance lines in 17 IUE
spectra (Hirsch~\cite{hirsch1998}). Recent developments in
stellar wind physics like in particular wind clumping (so far not
investigated in A-type supergiants) and possible effects of a weak magnetic field 
(Verdugo et al.~\cite{verdugo2002},~\cite{verdugo2005}) are not considered,
such that the derived wind parameters may be subject to revision in the future.
The need for a higher turbulent velocity than in the photospheric analysis, 
in order to reproduce the emission peak and remaining discrepancies in matching the 
absorption trough of H$\alpha$, are indicators that additional effects may need
to be considered in the modelling. However, this is beyond the scope of the
present paper. Note also that, at present, FASTWIND does not consider line formation for the metal 
ions relevant for A-type stars.

So far, several analyses of Deneb's atmosphere have been performed. 
The derived values of atmospheric parameters show a remarkable spread,
as summarised in Table \ref{parametersliterature}. Values for the effective temperature
range from 7635 to 10\,080\,K, with most studies indicating $T_{\rm eff}$ around
9000 to 10\,000\,K. Our determination indicates a low $T_{\rm eff}$, which is 
in excellent agreement with the comprehensive study of Aufdenberg et 
al.~(\cite{aufdenberg2002}).
The $\log g$ also shows a remarkable spread with our result
being lower than in most of the other studies. 
Finally, our value for the microturbulent velocity falls roughly in the
middle of the range discussed in the literature. In contrast to several
previous analyses, we derive a unique value for $\xi$ from all ions. Also, no
depth dependency of microturbulence is found in the photospheric analysis.

\subsection{Spectral energy distribution}
\label{sed}

\begin{figure}
   \centering
   \includegraphics[width=\columnwidth]{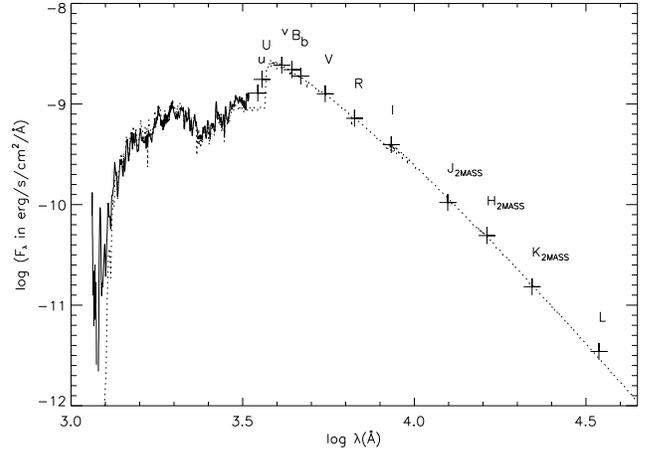}
\vspace{-0.7cm}
   \caption{Comparison of the ATLAS9 model flux (dotted line) with UV-spectra 
   from IUE (full line) and with measurements in various magnitude systems
   (Johnson, Str\"omgren, 2MASS). The theoretical flux is calculated for 
   our final parameters (Table \ref{allparameters}). The SEDs are normalised 
   in $V$. 
   }
   \label{fluxfit}
\end{figure}

We verify our results from the spectral line analysis of Deneb by comparing 
our model flux (from ATLAS9) with the observed spectral energy
distribution (SED).
Several sources of observational data are considered (see Sect.~\ref{observations}). 
We transform $m_1$, $b-y$, $c_1$ to Str\"omgren magnitudes $u$, $v$, $b$, $y$ 
assuming $V = y$. The various magnitudes are transformed into fluxes by adopting
zero points according to Heber et al.~(\cite{heber2002}, their Table~3). 
The observed fluxes are dereddened using a reddening law as described by
Cardelli et al.~(\cite{cardelli1989}), assuming a ratio of extinction 
to colour excess $A_V / E(B-V) = 3.1$. An overall good to excellent agreement between 
observed and computed SED is achieved, see Fig.~\ref{fluxfit}.

\subsection{Fundamental stellar parameters}
\label{fundparameters}

Finally, the spectral analysis allows us to determine the fundamental stellar parameters of Deneb, i.e. its luminosity $L$, mass $M$, and radius $R$. Adopting a distance modulus of 
$9.^{\hspace{-2pt}m}52$\,$\pm$\,$0.^{\hspace{-2pt}m}18$ from the affiliation of 
Deneb to the Cyg\,OB\,7 association (mean value of Humphreys~\cite{humphreys1978}), 
we derive the absolute visual magnitude and, using a bolometric correction from our 
model atmosphere computations, the absolute bolometric magnitude. The
resulting values for $L$, $R$, and $M$ are summarised in Table~\ref{allparameters}.
Excellent consistency is achieved with $R$ derived from the mean uniform-disk angular
diameter measurement of Deneb by Aufdenberg et al.~(\cite{aufdenberg2002})
assuming the same distance.

We can obtain an independent distance estimate by applying of the 
flux-weighted gravity-luminosity relationship (FGLR, Kudritzki et 
al.~\cite{kudritzki2003}). The FGLR allows
the absolute bolometric magnitude of BA-type supergiants to be derived from a measurement of 
$T_{\rm eff}$ and $\log g$.
We infer a distance modulus $(m_V-M_V)_0$ of $9.^{\hspace{-2pt}m}41\,\pm\,0.^{\hspace{-2pt}m}23$ 
with the FGLR, in good agreement with Humphreys~(\cite{humphreys1978}).

\begin{figure*}
   \sidecaption
   \centering
   \includegraphics[width=14.5cm]{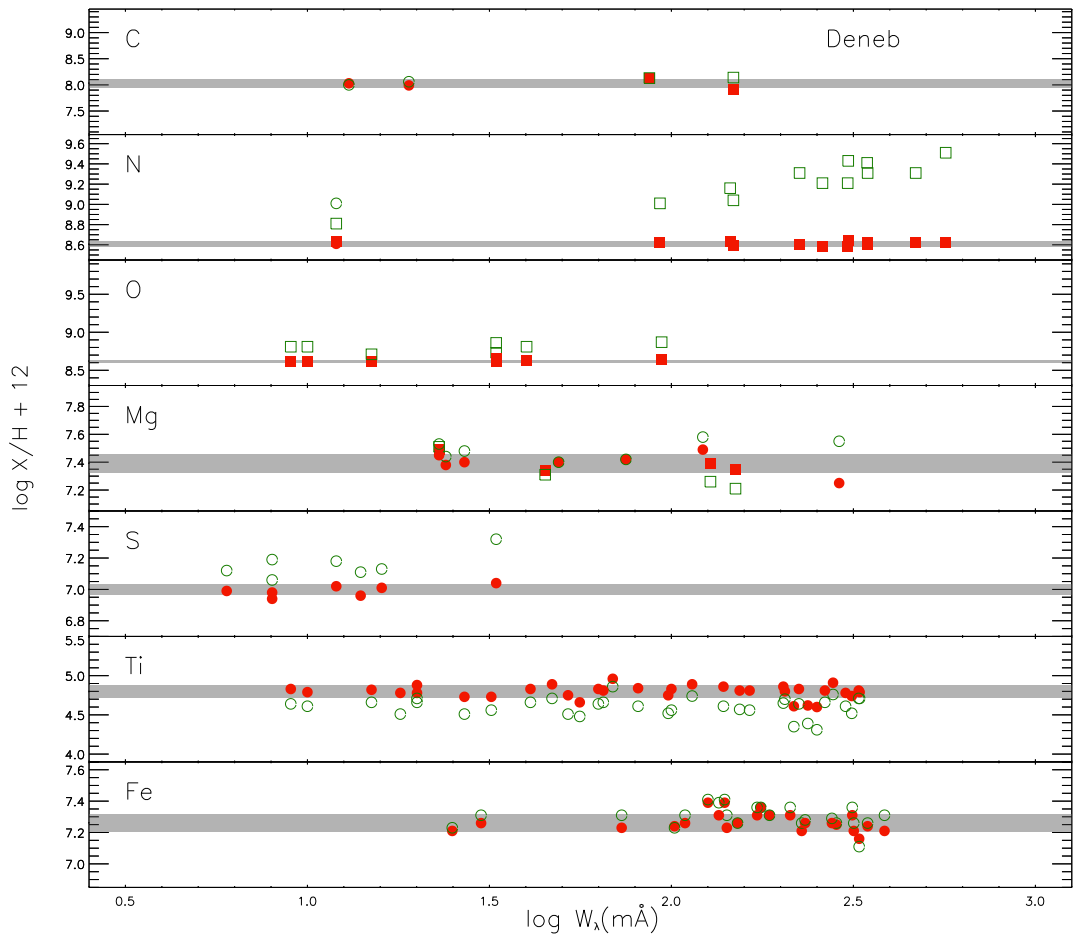}
\vspace{-0.7cm}
   \caption{An overview of non-LTE and LTE abundances for individual lines of all metals treated in non-LTE. The abundance derived from a single line is presented versus its equivalent width, if measurable. Filled red symbols indicate non-LTE abundances, empty green ones LTE abundances. Moreover, squares mean neutral species, circles single-ionized species. For each element, the non-LTE mean value with statistical uncertainties of all lines (including those analysed only by spectrum synthesis) is represented by grey bands. \vspace{1.0cm}}
   \label{abuvsw}
\end{figure*}

\begin{figure*}[ht!]
   \centering
   \includegraphics[width=\textwidth]{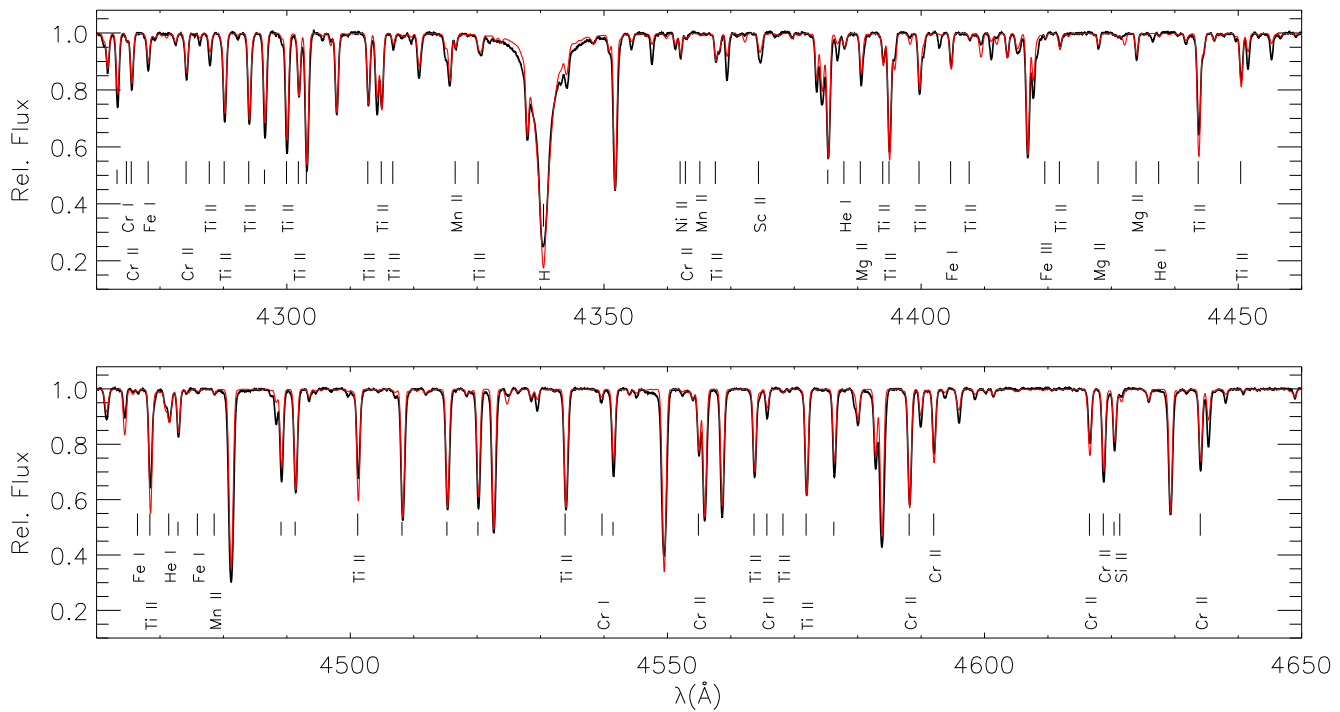}
\vspace{-0.7cm}
   \caption{Comparison of our spectrum synthesis (thin red line) with the high-resolution spectrum of Deneb (full black line). The spectral features used for our abundance determination are identified, short vertical marks designate \ion{Fe}{ii} lines.}
   \label{allelements}
\end{figure*}

\section{Abundances}
\label{abundances}

\subsection{Results of the present analysis}

We determine the abundances of several astrophysically interesting 
elements in non-LTE (He, C, N, O, Mg, S, Ti, and Fe)
and abundances of the remaining detectable species in LTE. The analysis is based on
detailed line-profile fits; equivalent widths are only measured as 
supplementary information. Details of the results from the analysis of individual 
lines can be found in the Appendix.
In the following, all abundance values are given in the usual logarithmic notation 
$\log \varepsilon({\rm el}) = \log \left( x_{\rm el} / x_{\rm H} \right) + 12$,
i.e. as number fractions $x_{\rm el}$ relative to hydrogen.

A comparison of the non-LTE and LTE abundances (i.e. considering and
neglecting non-LTE effects), where possible, is shown in Fig.~\ref{abuvsw}. 
The differences between non-LTE and LTE can be rather small in some cases, like carbon (usually below $\sim$0.05 dex).
However, in other cases like N and S, they can be larger (on the mean $\sim$0.6 and $\sim$0.2\,dex, respectively), demonstrating the improvements by the non-LTE
computations. Proper non-LTE calculations reduce the line-to-line scatter to only a few percent and remove systematic trends present in the LTE results. The S and Ti abundances show systematic 
differences in non-LTE and LTE. LTE abundances can be systematically
higher like in the case of S, whereas the situation for Ti is reversed. 
Mean non-LTE and LTE abundances of iron are in rather good agreement. The LTE analysis indicates larger statistical uncertainties 
in practically all cases.
It is usually assumed that non-LTE effects occur only in strong lines. Note, however, that even weak lines can show considerable 
departures from LTE as can be seen e.g. in the case of nitrogen.
Overall, similar trends for non-LTE effects, as in other Galactic BA-type supergiants (Przybilla et al.~\cite{przybilla2006}), are found.

Mean non-LTE and LTE abundances 
are summarised in Table~\ref{abundanceslist}. The statistical uncertainties of the non-LTE results 
are typically (much) smaller than for the LTE values. These small statistical 
uncertainties and the simultaneous establishment of all ionisation equilibria 
in non-LTE, i.e. a high degree of self-consistency, are characteristic for our analysis.
As a consequence, excellent agreement between our final synthetic
spectrum and observations is found (except for a few features), see Fig.~\ref{allelements} 
for an example.

\begin{table}
\caption{Mean non-LTE and LTE abundances/statistical uncertainties.}
\label{abundanceslist}
\centering
\begin{tabular}{l c c r}
\hline\hline
Ion  & $\log \varepsilon_{\rm non-LTE}$  & $\log \varepsilon_{\rm LTE}$ & \# lines \\
\hline
He\,{\sc i} &       11.09\,$\pm$\,    0.02 & -- &        8 \\
C\,{\sc i} &       8.02\,$\pm$\,     0.11 &       8.09\,$\pm$\,    0.08 &        3 \\
C\,{\sc ii} &       8.05\,$\pm$\,    0.07 &       8.06\,$\pm$\,    0.07 &        3 \\
N\,{\sc i} &       8.61\,$\pm$\,    0.03 &       9.19\,$\pm$\,     0.24 &       13 \\
N\,{\sc ii} &       8.61 &       9.01 &        1 \\
O\,{\sc i} &       8.62\,$\pm$\,    0.02 &       8.80\,$\pm$\,    0.07 &       15 \\
Ne\,{\sc i} & -- &       8.17\,$\pm$\,    0.02 &        4 \\
Na\,{\sc i} & -- &       7.02\,$\pm$\,     0.11 &        3 \\
Mg\,{\sc i} &       7.39\,$\pm$\,    0.07 &       7.32\,$\pm$\,     0.13 &        4 \\
Mg\,{\sc ii} &       7.39\,$\pm$\,    0.07 &       7.49\,$\pm$\,    0.07 &        8 \\
Al\,{\sc i} & -- &       5.98\,$\pm$\,    0.06 &        2 \\
Al\,{\sc ii} & -- &       6.11\,$\pm$\,     0.18 &        2 \\
Si\,{\sc ii} & -- &       7.67\,$\pm$\,    0.09 &        7 \\
S\,{\sc ii} &       7.00\,$\pm$\,    0.03 &       7.18\,$\pm$\,    0.09 &       10 \\
Ca\,{\sc ii} & -- &       5.79\,$\pm$\,     0.20 &        3 \\
Sc\,{\sc ii} & -- &       2.38\,$\pm$\,     0.11 &        2 \\
Ti\,{\sc ii} &       4.80\,$\pm$\,    0.08 &       4.60\,$\pm$\,     0.11 &       40 \\
V\,{\sc ii} & -- &       3.58\,$\pm$\,    0.03 &        8 \\
Cr\,{\sc i} & -- &       5.68\,$\pm$\,    0.08 &        5 \\
Cr\,{\sc ii} & -- &       5.63\,$\pm$\,    0.08 &       39 \\
Mn\,{\sc i} & -- &       5.40 &        1 \\
Mn\,{\sc ii} & -- &       5.39\,$\pm$\,    0.04 &       11 \\
Fe\,{\sc i} & -- &       7.20\,$\pm$\,    0.05 &       20 \\
Fe\,{\sc ii} &       7.26\,$\pm$\,    0.06 &       7.29\,$\pm$\,    0.07 &       28 \\
Fe\,{\sc iii} & -- &       7.28 &        1 \\
Ni\,{\sc ii} & -- &       6.19\,$\pm$\,    0.05 &        7 \\
Sr\,{\sc ii} & -- &       2.03\,$\pm$\,    0.04 &        2 \\
Ba\,{\sc ii} & -- &       2.03 &        1 \\
\hline
\end{tabular}
\end{table}

\begin{figure}[ht]
   \centering
   \includegraphics[width=\columnwidth]{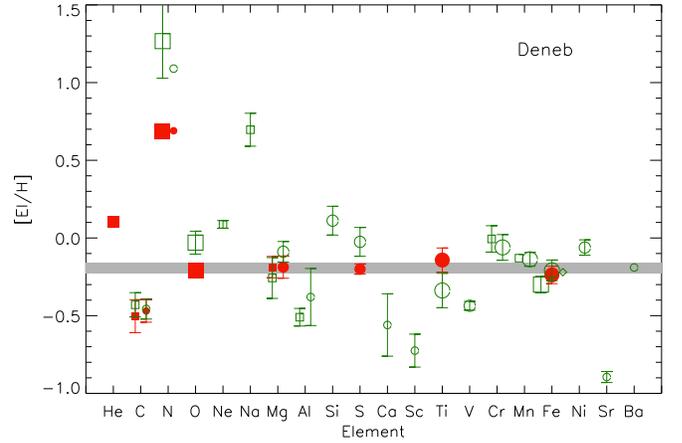}
\vspace{-0.7cm}
   \caption{Results from the elemental abundance analysis (relative to the solar composition, Grevesse \& Sauval \cite{grevesse1998}) for Deneb. Filled symbols denote non-LTE, open symbols LTE results. Boxes: neutral; circles: single-ionized; diamonds: double-ionized species. The symbol size codes the number of spectral lines analysed - small: 1 to 5; medium: 6 to 10; large: more than 10. Error bars represent 1$\sigma$-uncertainties from the line-to-line scatter. The grey-shaded area marks the deduced metallicity of Deneb within 1$\sigma$-errors. The non-LTE abundance analysis reveals abundances of $-$0.20\,$\pm$\,0.04\,dex (relative to solar composition) for all heavier elements in close agreement.}
   \label{abusolar}
\end{figure}

The results of our abundance analysis relative to the solar standard (Grevesse \& Sauval~\cite{grevesse1998}, meteoritic abundances favoured) are shown in Fig.~\ref{abusolar}. Combined with a slight overabundance of He (0.10\,dex), the strong overabundance of N (0.69\,dex), and the C-deficiency of 0.49\,dex indicate pronounced mixing with CN-processed matter from the stellar core. We obtain an N/C ratio of 4.44\,$\pm$\,0.84 and an N/O ratio of 0.86\,$\pm$\,0.06 (each by mass fraction). The implications of this abundance pattern on the evolutionary status of Deneb will be discussed in Sect.~\ref{evolutionarystatus}. Non-LTE abundances of oxygen and heavier elements show consistently sub-solar values of $-$0.20\,$\pm$\,0.04\,dex. A significant scatter around this mean value is found for the remaining elements, where only LTE abundances are available. The abundance pattern is analogous to the one in other Galactic BA-type supergiants (Przybilla et al.~\cite{przybilla2006}), but at significantly reduced metallicity. Future non-LTE investigations of those elements have to decide whether some of these deviations are intrinsic or whether they indicate systematic errors made by assuming of LTE for the line-formation calculations.

\subsection{Comparison with previous analyses}

Quantitative analyses of the chemical composition of Deneb are rare in the 
literature. The pioneering study of Groth~(\cite{groth1961}) derives an 
abundance pattern similar to ours, but it systematically finds larger abundances. 
Compared to our choice of the solar values, he finds an He-excess of 0.64\,dex,
a C-depletion by $-$0.33\,dex, and an N-/O-excess of 1.48 and 0.53\,dex,
respectively, as well as overabundances for other metals of typically $\sim$0.3$-$0.4\,dex. 
While outstanding for that time, the results of Groth have to be
critically reviewed in the context of improved model atmospheres and atomic data.

\begin{figure}
   \centering
   \includegraphics[width=\columnwidth]{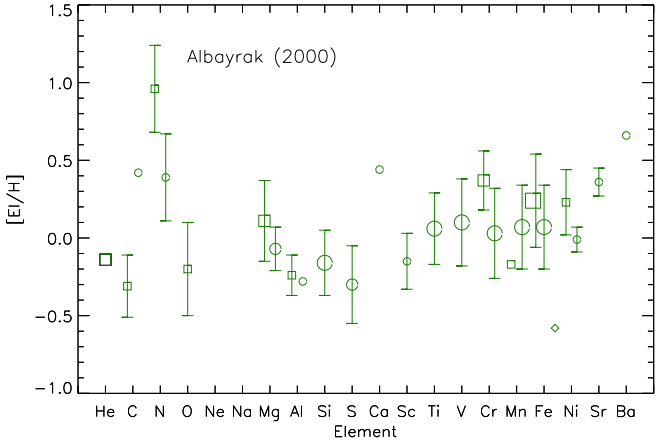}
\vspace{-0.7cm}
   \caption{Same as Fig. \ref{abusolar}, for the LTE analysis of Albayrak
   (\cite{albayrak2000}). Larger statistical uncertainties become apparent than in our non-LTE, and even our LTE analysis and systematic effects occur. Striking features are the He-\emph{deficiency} and discrepancies in some ionisation equilibria (e.g. \ion{C}{i/ii} and \ion{Fe}{i/ii} vs. \ion{Fe}{iii}).}
   \label{abusolar_albayrak}
\end{figure}

The most recent abundance analysis of Deneb is provided by
Albayrak~(\cite{albayrak2000}). His results are summarised in Fig.~\ref{abusolar_albayrak}. 
The pure LTE analysis finds striking features like an He-\emph{deficiency} and 
large differences in abundances from different ions of an element, i.e. a failure 
to establish some of the ionisation equilibria. Albayrak's
analysis indicates a roughly solar composition on the mean,
with the abundances of the heavier elements tending to be slightly supersolar. 
However, the abundance determination gives large statistical
uncertainties and shows a wide spread around the solar standard, which 
in view of our findings can be interpreted as systematic uncertainties. 
These may be the consequence of a combination of systematically biased
stellar parameters -- in particular the higher $T_{\rm eff}$ (see
Table~\ref{parametersliterature}) will result in
larger abundances for most of the ionic species -- and the neglect of non-LTE effects.

\begin{figure}
   \centering
   \includegraphics[width=\columnwidth]{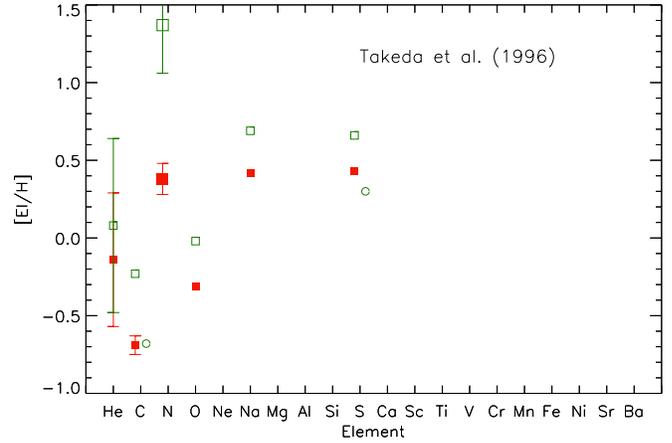}
\vspace{-0.7cm}
   \caption{Same as Fig. \ref{abusolar}, for the results of Takeda et al.~(1996). Non-LTE calculations were made only for neutral species. Note that He tends to be depleted, while the heavier elements indicate super-solar abundances, unlike our results.}
   \label{abusolar_takeda}
\end{figure}

Takeda et al.~(\cite{takeda1996}) also provide an analysis of 
elemental abundances in Deneb, partially in non-LTE (Fig.~\ref{abusolar_takeda}). 
They derive a moderate N-excess by 0.3\,dex and a significant C-deficiency of 
about 0.5\,dex relative to the solar standard, 
in moderate agreement with our results. Helium tends to be \emph{depleted} 
by 0.14\,dex; however, an uncertainty of 0.43\,dex renders any conclusion difficult. The heavy elements (Na and S) show 
super-solar abundances, while oxygen is significantly sub-solar. 
The higher $T_{\rm eff}$ adopted in the work of Takeda et al.~(see
Table~\ref{parametersliterature}) will result in systematically larger
abundances like in the case of Albayrak~(\cite{albayrak2000}).

The direct comparison of Fig.~\ref{abusolar} with
Figs.~\ref{abusolar_albayrak} and~\ref{abusolar_takeda} demonstrates the
drastic (note the logarithmic scale) reduction of statistical {\em and} 
systematic uncertainties in the present work. The existence of abundance patterns like those
derived in the two other abundance studies are
difficult to explain in terms of a) massive star evolution: mixing with
CN-cycled matter results in helium enhancement; and b) galactochemical
evolution: massive stars in the solar neighbourhood have slightly
sub-solar, not (markedly) super-solar, metallicities (see e.g. Table~4 of
Przybilla et al.~\cite{przybilla2006}). The present work leads to more
conservative conclusions, as we find a uniform pattern consistent with a metallicity of $-$0.2\,dex.

\section{Near-IR spectroscopy}
\label{irspectroscopy}

Aufdenberg et al.~(\cite{aufdenberg2002}) provide an analysis of the (near-)IR spectrum of Deneb reporting some discrepancies in particular concerning the Pfund series of hydrogen that was fitted for the first time. Thus, we also test our modelling in the same spectral regions, which are scarcely considered in quantitative analyses of A-type supergiants.
The high-quality near-IR spectra of Fullerton \& Najarro~(\cite{fullerton1998}) (a comprehensive data
set like this one is unavailable for any other A-type supergiant) are a key element in our analysis. 

Consistency in the
analysis of visual and near-IR spectra is not easily achieved because of the
amplification of non-LTE effects in the Rayleigh-Jeans tail of the spectral
energy distribution. This results in a high sensitivity of the
spectral lines not only to details in the atmospheric structure but also to the
details of the model atoms used in the statistical eqilibrium calculations. For hydrogen, this requires reliable cross-sections for excitation by electron impact in particular (Przybilla \& Butler~\cite{przybilla2004}), in addition to the accuractely known atomic data for radiative transitions.

\begin{figure*}
\centering
	\includegraphics[width=1.0\textwidth]{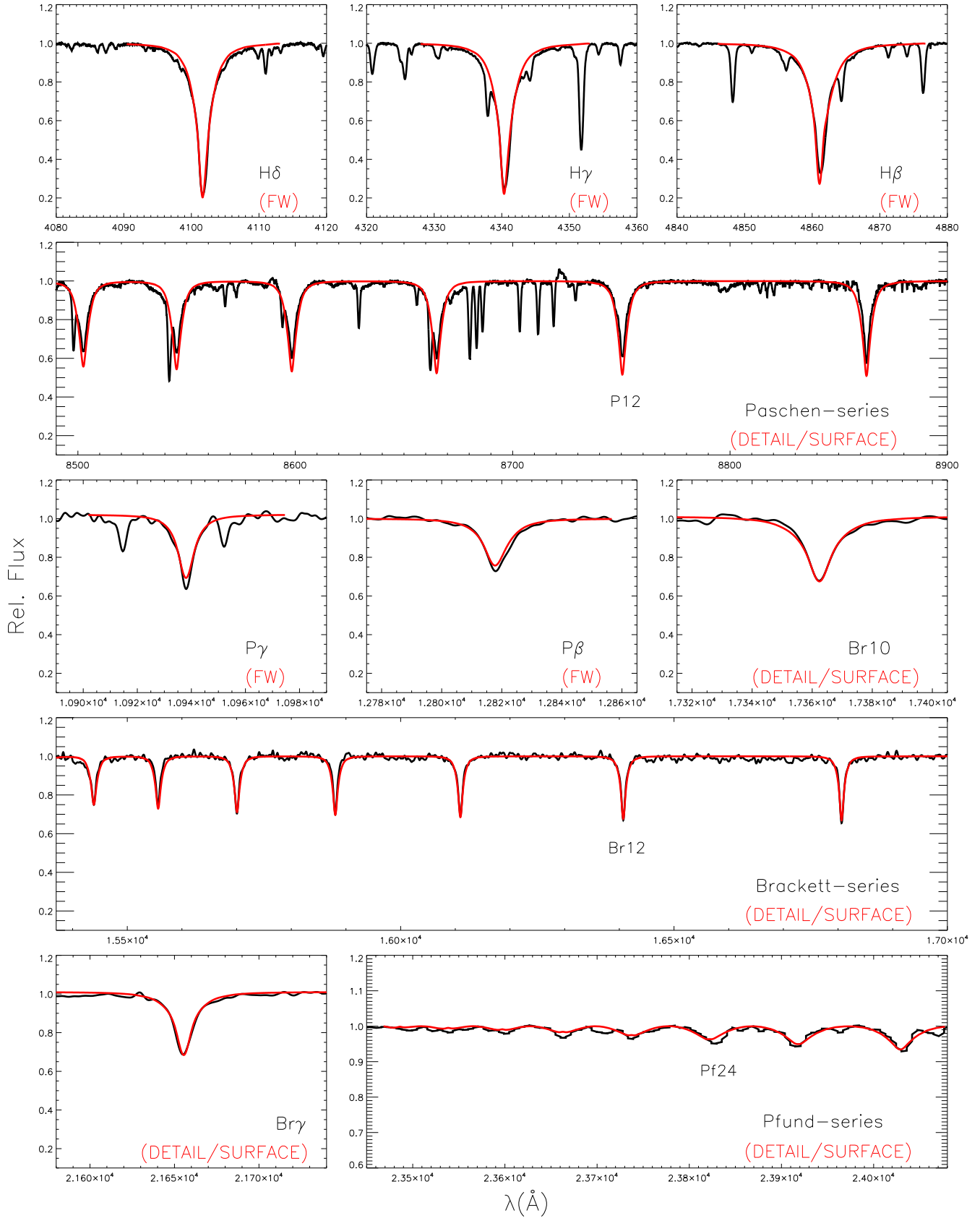}
\vspace{-0.6cm}
	\caption{Modelling (red) of the hydrogen lines in the visual and the near-IR (black curve). The synthetic spectra are calculated with our hybrid non-LTE technique using DETAIL/SURFACE (photospheric lines) or FASTWIND (FW, wind-affected lines), as indicated. Except for the Pfund series, all panels show the same range in relative flux. Some lines, such as P$\beta$ and H$\beta$, are noticeably affected by the stellar wind.}
	\label{infrared_overview}
\end{figure*}

Good to excellent agreement between spectrum synthesis with DETAIL/SURFACE
(photospheric lines) and FASTWIND (wind-affected lines) and the observations is
achieved comprising multiple features of the Paschen, Brackett, and Pfund series in addition to the traditionally utilised Balmer lines, as summarised in Fig.~\ref{infrared_overview}. Neither LTE nor non-LTE modelling based on collisional data derived from classical approximation formulae succeeds in obtaining this self-consistency (see also Przybilla \&
Butler~\cite{przybilla2004} for a detailed discussion). This resolves the
discrepancies reported by Aufdenberg et al.~(\cite{aufdenberg2002}).

\section{Evolutionary status}
\label{evolutionarystatus}

Massive stars are important drivers of the chemodynamic evolution of galaxies 
and their interstellar medium. A detailed investigation of these topics requires
quantitative understanding of the evolution of massive stars in general and
of the advanced stages in their lives in particular. In the
following we want to discuss our results for the benchmark A-type supergiant Deneb in the
context of the stellar evolution models available at present.

\begin{figure}
   \centering
   \includegraphics[width=\columnwidth]{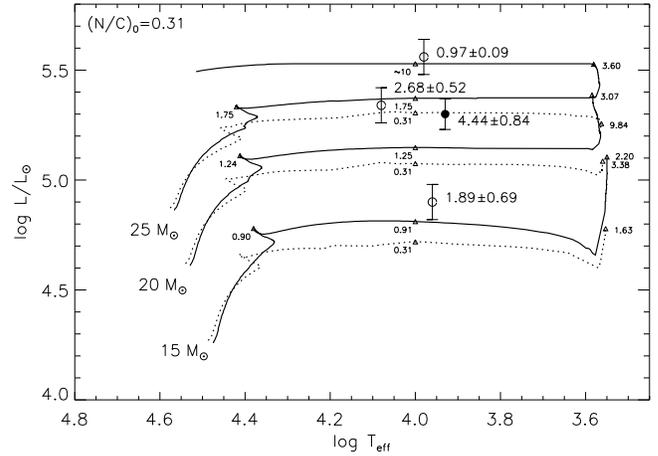}
\vspace{-0.7cm}
   \caption{Stellar evolution tracks (Meynet \& Maeder \cite{meynet2003}) in the HRD for different ZAMS masses of stars with an initial rotational velocity of 300\,km\,s$^{-1}$ (full lines) or 0\,km\,s$^{-1}$ (dotted lines) and solar metallicity until the end of central helium burning. Observed N/C ratios (large numbers) are displayed for a few objects: Deneb (dot) and three stars from Przybilla et al.~(\cite{przybilla2006}, open circles). Small numbers along the curves indicate the progression of N/C-ratios (mass fractions) in the course of evolution, assuming an initial value of 0.31 (solar). The N/C ratios at the red supergiant stage mark the situation at the end of central helium burning, with the exception of the 25\,$M_\odot$ model with rotation which enters the Wolf-Rayet phase near the end of central helium burning.}
   \label{evolution}
\end{figure}

A Hertzsprung-Russell diagram (HRD) is shown in Fig.~\ref{evolution}. Evolution
tracks for rotating and non-rotating stars according to Meynet \&
Maeder~(\cite{meynet2003}) are displayed. Starting with an initial N/C ratio
(by mass) of 0.31, the stars may experience rotationally-induced mixing of the 
atmospheric layers with nuclear-processed material. Predicted N/C ratios at the end of
H-burning, in the blue supergiant stage (at $T_{\rm
eff}$\,$=$\,10\,000\,K), and in the red supergiant stage after the first
dredge-up are also displayed. Deneb falls between the tracks for
zero-age main-sequence (ZAMS) masses of 20 and 25\,$M_\odot$ and shows a much more
pronounced N/C ratio than predicted. The value is also much higher than in 
similar supergiants analysed with the same method 
(Przybilla et al.~\cite{przybilla2006}). This leaves us with two possible scenarios for
Deneb's state of evolution.

In the first scenario, Deneb started its life as a late O-type star with 
$M^{\rm ZAMS}$\,$=$\,23\,$\pm$\,2\,M$_\odot$ on the main sequence and is
currently evolving to the red supergiant stage. The second scenario 
implies a more advanced stage of an $M^{\rm ZAMS}$\,$\simeq$\,20\,$\pm$\,2\,M$_\odot$
star after the red supergiant phase, which may also lead it through the blue
supergiant regime (Hirschi et al.~\cite{hirschi2004}). In this case
Deneb's atmosphere should expose abundance ratios for the
light elements typical of the first dredge-up phase.

The two scenarios may be distinguished by a comparison of spectroscopic and 
evolutionary masses and by the light element abundances. The first scenario
implies an evolutionary mass of $M^{\rm evol}$\,$\simeq$\,18\,$\pm$\,2\,M$_\odot$, 
in excellent agreement with the derived 
$M^{\rm spec}$\,$\simeq$\,19\,$\pm$\,4\,M$_\odot$, whereas the predicted
mass in the second
scenario, $M^{\rm evol}$\,$\simeq$\,11\,M$_\odot$, is inconsistent with
$M^{\rm spec}$. At first glance, the observed high N/C ratio agrees better with a first dredge-up scenario, though helium would be underabundant.
However, the predicted surface abundances depend on the assumed initial rotational velocity of the star. The tracks
follow the evolution of a massive star with typical initial angular momentum, i.e. with a fixed rotational velocity of 300\,km\,s$^{-1}$. If the star rotates even faster, the He abundance and the N/C ratio
may be higher. This may even be enhanced further by
the interplay of rotation with a magnetic field 
(Maeder \& Meynet~\cite{maeder2005}). 

We conclude that Deneb is a star evolving from the main sequence to the red
supergiant stage. Because of its pronounced mixing signature with
nuclear-processed matter, its main-sequence progenitor was probably an
initially fast rotator.

\section{Summary and conclusions}
\label{summaryandconclusions}

A detailed and comprehensive model atmosphere analysis of the prototype
A-type supergiant Deneb was performed. We derived basic atmospheric
parameters, elemental abundances, and fundamental stellar parameters
as summarised in Tables~\ref{allparameters} and~\ref{abundanceslist} using a
hybrid non-LTE spectrum
synthesis technique, as well as line-blanketed hydrodynamic non-LTE
models.
High consistency in the results was achieved, bringing all indicators
simultaneously into match: multiple metal ionisation equilibria,
Stark-broadened
hydrogen lines from the Balmer to Pfund series, and the spectral energy
distribution from the UV to the near-IR. The results were
obtained with high internal accuracy.

Our detailed non-LTE abundance analysis reveals an abundance pattern typical of the mixing of the atmosphere with CN-processed matter from the stellar core, i.e. He- and N-enrichment combined with C-depletion. Non-LTE abundances of the heavier elements consistently show a value of $\sim$0.20\,dex below the solar standard, exhibiting none of the peculiar patterns reported in earlier analyses.

The evolutionary status of Deneb was constrained from a comparison of the
observed properties with stellar evolution models. The spectroscopic mass, the
surface He abundance, and a high N/C ratio are indicative of the evolution
of an initially fast-rotating main-sequence star towards the red supergiant
stage.

\begin{acknowledgements}

We would like to thank U.~Heber for his support and interest in the project, as well as for many useful comments on the manuscript. We thank A.W. Fullerton for providing the high-quality infrared spectra in the Jw-, H-, and K-band. We also thank A.F. Gulliver for making available an electronic version of a spectral atlas of Deneb. We are grateful to K. Butler for making DETAIL and SURFACE available and to J. Puls for making FASTWIND available. F. Schiller would like to acknowledge the financial support of the "Studienstiftung des deutschen Volkes" which makes this project feasible throughout the Ph.D. phase.

\end{acknowledgements}

\appendix

\Online

\noindent\large{\textbf{Appendix A: Spectral line analysis}}

\vspace{5mm}

\noindent In this appendix we provide details on our spectral line analysis,
as a basis for further interpretation. Table A1 summarises our line data and
the results from the abundance analysis for Deneb. The first columns give
the transition wavelength $\lambda$ (in {\AA}), excitation energy of the
lower level $\chi$ (in eV), adopted oscillator strength $\log gf$, an
accuracy indicator, and the source of the $gf$ value. Then, for each line the
measured equivalent width $W_\lambda$ (in m{\AA}) is tabulated, followed by
the derived abundance $\log \varepsilon = \log(\mbox{X/H}) + 12$. In cases
with an entry for the non-LTE abundance correction $\Delta \log \varepsilon
= \log \varepsilon_{\mbox{non-LTE}} - \log \varepsilon_{\mbox{LTE}}$ this 
denotes the non-LTE abundance, otherwise the LTE abundance. The equivalent widths 
have been measured by direct integration over the spectral lines. In some 
cases metal lines are situated in the wings of the Balmer lines. Then their 
equivalent width is measured with respect to the \emph{local} continuum, 
indicated by 'S($W_\lambda$)'. Abundances are determined from a best match of 
the spectrum synthesis to the observed line \emph{profiles} and not the 
equivalent widths. Spectrum synthesis also allows blended features to be used 
for abundances determinations, where a $W_\lambda$ measurement is hampered. 
These cases are marked by a single `S' in Table A1. For \ion{He}{i}, only non-LTE 
abundances are derived because of the potential of this major atmospheric 
constituent to change atmospheric structure and thus the stellar parameter 
determination.

\clearpage

\begin{multicols}{2}

\begin{table*}
\setlength{\tabcolsep}{.0625cm}
\textbf{Table A1} Spectral line analysis\\\label{taba1}
\begin{tabular}{lrrlrlrrr}
\hline\hline
\rule[-2mm]{0mm}{1mm}{}\\[-4.mm]
$\lambda\,$({\AA}) & $\chi\,(\mbox{eV})$ & $\log gf$ & Acc. & Src. & & 
$W_{\lambda}$(m{\AA}) & $\log \varepsilon$ &
$\Delta \log \varepsilon$ \\[.8mm]
\hline
He\,{\sc i}:\\
4026.18 & 20.96 & $-$2.63 & A & WSG & & 22 & 11.09 & {\ldots}\\
4026.19 & 20.96 & $-$0.63 & A & WSG\\
4026.20 & 20.96 & $-$0.85 & A & WSG\\
4026.36 & 20.96 & $-$1.32 & A & WSG\\
4120.81 & 20.96 & $-$1.74 & B & WSG & & 34 & 11.12 & {\ldots}\\
4120.82 & 20.96 & $-$1.96 & B & WSG\\
4120.99 & 20.96 & $-$2.44 & B & WSG\\
4387.93 & 21.22 & $-$0.88 & A & WSG & & S & 11.13 & {\ldots}\\
4437.55 & 21.22 & $-$2.03 & B & WSG & & 5 & 11.09 & {\ldots}\\
4471.47 & 20.96 & $-$0.20 & A & WSG & & S & 11.07 & {\ldots}\\
4471.49 & 20.96 & $-$0.42 & A & WSG\\
4471.68 & 20.96 & $-$0.90 & A & WSG\\
4713.14 & 20.96 & $-$1.23 & B & WSG & & 30 & 11.10 & {\ldots}\\
4713.16 & 20.96 & $-$1.45 & B & WSG\\
4713.38 & 20.96 & $-$1.93 & B & WSG\\
5015.68 & 20.62 & $-$0.82 & AA & WSG & & S & 11.09 & \ldots \\
5875.60 & 20.96 & $-$1.52 & A & WSG & & 101 & 11.06 & \ldots \\
5875.61 & 20.96 & 0.48 & A & WSG\\
5875.63 & 20.96 & $-$0.34 & A & WSG\\
5875.64 & 20.96 & 0.14 & A & WSG\\
5875.97 & 20.96 & $-$0.22 & A & WSG\\[1.5mm]
C\,{\sc i}:\\
4771.74 & 7.49 & $-$1.87 & C & WFD &  & 87 & 8.13 & 0.00\\
5052.17 & 7.68 & $-$1.45 & B & WFD &  & S & 8.00 & 0.00\\
9088.51 & 7.48 & $-$0.43 & B & WFD &  & 148 & 7.92 & $-$0.22\\[1.5mm]
C\,{\sc ii}:\\
4267.00 & 18.05 & 0.56 & C+ & WFD &  & 19 & 7.99 & $-$0.07\\
4267.26 & 18.05 & 0.74 & C+ & WFD\\
6578.05 & 14.45 & $-$0.03 & B & WFD &  & S(13) & 8.03 & 0.03\\
6582.88 & 14.45 & $-$0.33 & B & WFD &  & S & 8.13 & 0.00\\[1.5mm]
N\,{\sc i}:\\
6008.47 & 11.60 & $-$1.11 & C+ & WFD & & 12 & 8.63 & $-$0.18 \\
7423.64 & 10.33 & $-$0.71 & B+ & WFD & & 145 & 8.63 & $-$0.53 \\
7442.30 & 10.33 & $-$0.38 & B+ & WFD & & 225 & 8.60 & $-$0.71 \\
7468.31 & 10.34 & $-$0.19 & B+ & WFD & & 306 & 8.64 & $-$0.79 \\
7898.98 & 12.36 & 0.02 & C & WFD & & S & 8.55 & $-$0.16 \\
7899.28 & 12.36 & $-$0.91 & C & WFD\\
8567.74 & 10.68 & $-$0.66 & B & WFD & & S(148) & 8.59 & $-$0.45 \\
8680.28 & 10.34 & 0.35 & B+ & WFD & & S(567) & 8.62 & $-$0.89 \\
8683.40 & 10.33 & 0.09 & B+ & WFD & & S(469) & 8.62 & $-$0.69 \\
8686.15 & 10.33 & $-$0.31 & B+ & WFD & & S(347) & 8.62 & $-$0.79 \\
8703.25 & 10.33 & $-$0.32 & B+ & WFD & & 305 & 8.58 & $-$0.63 \\
8711.70 & 10.33 & $-$0.23 & B+ & WFD & & 346 & 8.60 & $-$0.71 \\
8718.84 & 10.34 & $-$0.34 & B+ & WFD & & 260 & 8.58 & $-$0.63 \\
8728.90 & 10.33 & $-$1.07 & B+ & WFD & & S(93) & 8.62 &$-$0.39 \\[1.5mm]
N\,{\sc ii}:\\
3995.00 & 18.50 & 0.21 & B & WFD & & 12 & 8.61 & $-$0.40 \\[1.5mm]
O\,{\sc i}:\\
3947.29 &  9.15 & $-2.10$ & B & WFD & & 33 & 8.61 & $-$0.12 \\
3947.48 &  9.15 & $-2.24$ & B & WFD \\
3947.59 &  9.15 & $-2.47$ & B & WFD \\
4654.12 & 10.74 & $-2.16$ & C+ & WFD & & S & 8.61 & $-$0.10 \\
4654.56 & 10.74 & $-1.93$ & C+ & WFD\\
4772.45 & 10.74 & $-1.92$ & C+ & WFD & & S & 8.61 & $-$0.10 \\
4772.91 & 10.74 & $-1.70$ & C+ & WFD\\
4773.75 & 10.74 & $-1.55$ & C+ & WFD & & S & 8.61 & $-$0.10 \\
\hline
\end{tabular}
\end{table*}

\begin{table*}
\noindent {\bf Table A1} (cont.) \\[1mm]
\setlength{\tabcolsep}{.0625cm}
\begin{tabular}{lrrlrlrrr}
\hline\hline
\rule[-2mm]{0mm}{1mm}{}\\[-4.mm]
$\lambda\,$({\AA}) & $\chi\,(\mbox{eV})$ & $\log gf$ & Acc. & Src. & & 
$W_{\lambda}$(m{\AA}) & $\log \varepsilon$ &
$\Delta \log \varepsilon$ \\[.8mm]
\hline
O\,{\sc i}:\\
4967.38 & 10.74 & $-1.63$ & C+ & WFD & & 15 & 8.61 & $-$0.10 \\
4967.88 & 10.74 & $-1.41$ & C+ & WFD \\
5329.10 & 10.74 & $-1.24$ & C+ & WFD & & 40 & 8.63 & $-$0.18 \\
5329.68 & 10.74 & $-1.02$ & C+ & WFD \\
5330.73 & 10.74 & $-0.87$ & C+ & WFD & & 33 & 8.66 & $-$0.20 \\
5435.18 & 10.74 & $-1.78$ & C+ & WFD & & 10 & 8.61 & $-$0.20 \\%
5435.77 & 10.74 & $-1.54$ & C+ & WFD \\%
5436.86 & 10.74 & $-1.39$ & C+ & WFD & & 9 & 8.61 & $-$0.20 \\%
6046.23 & 10.99 & $-1.76$ & C+ & WFD & & S & 8.61 & $-$0.20 \\
6046.44 & 10.99 & $-1.54$ & C+ & WFD \\
6046.49 & 10.99 & $-2.24$ & C+ & WFD \\
6155.96 & 10.74 & $-1.36$ & B+ & WFD & & S & 8.63 & $-$0.18 \\
6155.97 & 10.74 & $-1.01$ & B+ & WFD \\
6155.99 & 10.74 & $-1.12$ & B+ & WFD \\
6156.74 & 10.74 & $-1.49$ & B+ & WFD \\
6156.76 & 10.74 & $-0.90$ & B+ & WFD \\
6156.78 & 10.74 & $-0.69$ & B+ & WFD \\
6158.15 & 10.74 & $-1.84$ & B+ & WFD & & 94 & 8.64 & $-$0.23 \\
6158.17 & 10.74 & $-1.00$ & B+ & WFD \\
6158.19 & 10.74 & $-0.41$ & B+ & WFD \\
6453.60 & 10.74 & $-1.29$ & C+ & WFD & & S & 8.61 & $-$0.35 \\
6454.44 & 10.74 & $-1.07$ & C+ & WFD \\
7001.90 & 10.99 & $-1.49$ & B & WFD & & S & 8.61 & $-$0.25 \\
7001.92 & 10.99 & $-1.01$ & B & WFD \\
7002.17 & 10.99 & $-2.66$ & B & WFD \\
7002.20 & 10.99 & $-1.49$ & B & WFD \\
7002.23 & 10.99 & $-0.74$ & B & WFD \\
7002.25 & 10.99 & $-1.36$ & B & WFD \\
7254.15 & 10.99 & $-1.27$ & C+ & WFD & & S & 8.66 & $-$0.17 \\
7254.45 & 10.99 & $-1.05$ & C+ & WFD \\
7254.53 & 10.99 & $-1.74$ & C+ & WFD \\[1.5mm]
Ne\,{\sc i}:\\
5852.49 & 16.85 & $-$0.49 & B & S & & 5 & 8.14 & {\ldots} \\
6143.06 & 16.62 & $-$0.10 & B & S & & 7 & 8.16 & {\ldots} \\
6266.50 & 16.72 & $-$0.37 & B & S & & 5 & 8.17 & {\ldots} \\
6402.20 & 16.62 & 0.33 & B & S & & 15 & 8.20 & {\ldots} \\[1.5mm]
Na\,{\sc i}:\\
4982.81 & 2.10 & $-$0.91 & D & WSM & & S & 7.13 & {\ldots} \\%
5682.63 & 2.10 & $-$0.70 & C & WSM & & S & 7.00 & {\ldots} \\%
8194.79 & 2.10 & $-$0.44 & D & WSM & & 60 & 6.92 & {\ldots} \\%
8194.82 & 2.10 & 0.51 & C & WSM \\[1.5mm]%
Mg\,{\sc i}:\\
4702.99 & 4.35 & $-$0.42 & C+ & BMZ & & 23 & 7.49 & $-$0.02 \\
5172.68 & 2.71 & $-$0.38 & B & WSM & & 128 & 7.39 & $+$0.13 \\
5183.60 & 2.72 & $-$0.16 & B & WSM & & 150 & 7.35 & $+$0.14 \\
8806.76 & 4.35 & $-$0.16 & C+ & BMZ & & 45 & 7.34 & $+$0.03 \\[1.5mm]
Mg\,{\sc ii}:\\
4390.51 & 10.00 & $-$1.71 & D & WSM & & 122 & 7.49 & $-$0.09 \\
4390.57 & 10.00 & $-$0.53 & D & WSM \\
4427.99 & 10.00 & $-$1.20 & C+ & WSM & & 27 & 7.40 & $-$0.08 \\
4433.99 & 10.00 & $-$0.90 & C+ & WSM & & 49 & 7.40 & 0.00 \\
4739.71 & 11.57 & $-$0.66 & C+ & T & & 24 & 7.38 & -0.06 \\
4739.71 & 11.57 & $-$0.77 & C+ & T \\
4851.08 & 11.63 & $-$0.42 & C & CA & & S(23) & 7.45 & -0.08 \\
6545.97 & 11.63 & 0.41 & C & CA & & S(75) & 7.42 & 0.00 \\
7877.05 & 10.00 & 0.39 & C+ & WSM & & S & 7.35 & $-$0.18 \\
7896.04 & 10.00 & $-$0.30 & C+ & WSM & & 289 & 7.25 & $-$0.30 \\
7896.37 & 10.00 & 0.65 & C+ & WSM \\[1.5mm]
\hline
\end{tabular}
\end{table*}

\begin{table*}
\noindent {\bf Table A1} (cont.) \\[1mm]
\setlength{\tabcolsep}{.0625cm}
\begin{tabular}{lrrlrlrrr}
\hline\hline
\rule[-2mm]{0mm}{1mm}{}\\[-4.mm]
$\lambda\,$(\AA) & $\chi\,(\mbox{eV})$ & $\log gf$ & Acc. & Src. & & 
$W_{\lambda}$(m\AA) & $\log \varepsilon$ &
$\Delta \log \varepsilon$ \\[.8mm]
\hline
Al\,{\sc i}:\\
3944.01 & 0.00 & $-$0.64 & C+ & WSM & & 64 & 6.02 & {\ldots} \\
3961.52 & 0.01 & $-$0.34 & C+ & WSM & & 80 & 5.94 & {\ldots} \\[1.5mm]
Al\,{\sc ii}:\\
4663.05 & 10.60 & $-$0.29 & C & WSM & & S & 6.24 & {\ldots} \\
5593.30 & 13.26 & 0.41 & D & WSM & & S & 5.98 & {\ldots} \\[1.5mm]
Si\,{\sc ii}:\\
4128.07 & 9.84 & 0.31 & C & WSM & & S & 7.73 & {\ldots} \\%
4130.89 & 9.84 & 0.49 & C & WSM & & 264 & 7.71 & {\ldots} \\%
4621.42 & 12.53 & $-$0.54 & D & WSM & & 18 & 7.47 & {\ldots} \\
4621.70 & 12.53 & $-$1.68 & D & WSM\\
4621.72 & 12.53 & $-$0.39 & D & WSM\\
5041.02 & 10.07 & 0.17 & X & MEL & & 201 & 7.70 & {\ldots} \\
5688.81 & 14.16 & 0.16 & X & MEL & & S & 7.72 & {\ldots} \\
5957.56 & 10.07 & $-$0.35 & D & WSM & & S & 7.65 & {\ldots} \\
5978.93 & 10.07 & $-$0.06 & D & WSM & & 143 & 7.72 & {\ldots} \\[1.5mm]
S\,{\sc ii}:\\
4142.26 & 15.85 & 0.24 & D$-$ & WSM & & S & 7.04 & $-$0.16 \\%
4153.07 & 15.90 & 0.62 & D$-$ & WSM & & 12 & 7.02 & $-$0.16 \\
4162.67 & 15.94 & 0.78 & D$-$ & WSM & & 14 & 6.96 & $-$0.15 \\
4716.27 & 13.62 & $-$0.41 & D & WSM & & 8 & 6.94 & $-$0.12 \\
4815.55 & 13.67 & 0.09 & D & WSM & & 16 & 7.01 & $-$0.12 \\
4917.20 & 14.00 & $-$0.32 & D & WSM & & 6 & 6.99 & $-$0.13 \\%
5320.72 & 15.07 & 0.50 & D & WSM & & 8 & 6.98 & $-$0.21 \\
5428.66 & 13.58 & $-$0.13 & D & WSM & & S & 7.00 & $-$0.10 \\
5453.86 & 13.67 & 0.48 & D & WSM & & 33 & 7.04 & $-$0.28 \\
5660.00 & 13.68 & $-$0.05 & D & WSM & & S & 7.02 & $-$0.32 \\[1.5mm]
Ca\,{\sc ii}:\\
4799.97 & 8.44 & $-$0.42 & C & SA & & S & 6.02 & {\ldots} \\
8248.80 & 7.51 & 0.57 & C$-$ & WSM & & 50 & 5.65 & {\ldots} \\
8912.07 & 7.05 & 0.57 & X & KB & & 103 & 5.70 & {\ldots} \\[1.5mm]
Sc\,{\sc ii}:\\
4246.82 & 0.32 & 0.28 & D & KB & & 143 & 2.30 & {\ldots} \\
4374.46 & $-$0.44 & $-$0.64 & D & KB & & S & 2.45 & {\ldots} \\[1.5mm]
Ti\,{\sc ii}:\\
3900.56 & 1.13 & $-$0.45 & D & MFW & & 329 & 4.79 & +0.08 \\
3913.48 & 1.12 & $-$0.53 & D & MFW & & 301 & 4.78 & +0.17 \\
3987.60 & 0.61 & $-$2.73 & D & MFW & & 20 & 4.78 & +0.12 \\%
4025.13 & 1.13 & $-$0.45 & D & MFW & & S & 4.76 & +0.25 \\
4028.36 & 1.89 & $-$1.00 & D & MFW & & 114 & 4.89 & +0.15 \\
4287.88 & 1.08 & $-$2.02 & D$-$ & MFW & & 69 & 4.96 & +0.10 \\
4290.22 & 1.16 & $-$1.12 & D$-$ & MFW & & 203 & 4.86 & +0.21 \\
4290.35 & 2.06 & $-$1.53 & X & KB\\
4294.09 & 1.08 & $-$1.11 & D$-$ & MFW & & 205 & 4.80 & +0.10 \\
4300.06 & 1.18 & $-$0.77 & D$-$ & MFW & & 278 & 4.91 & +0.15 \\
4301.92 & 1.16 & $-$1.16 & D$-$ & MFW & & 164 & 4.81 & +0.25 \\
4312.87 & 1.18 & $-$1.16 & D$-$ & MFW & & 154 & 4.81 & +0.24 \\
4314.97 & 1.16 & $-$1.13 & D$-$ & MFW & & S & 4.81 & +0.24 \\
4316.80 & 2.05 & $-$1.42 & D & MFW & & 65 & 4.81 & +0.15 \\%
4330.24 & 2.04 & $-$1.51 & D & MFW & & S(63) & 4.83 & +0.19 \\
4330.72 & 1.18 & $-$2.04 & D$-$ & MFW \\
4367.66 & 2.59 & $-$0.72 & X & KB & & S & 4.81 & +0.25 \\%
4394.02 & 1.22 & $-$1.59 & D$-$ & MFW & & 52 & 4.75 & +0.24 \\
4395.00 & 1.08 & $-$0.66 & D$-$ & MFW & & 313 & 4.74 & +0.22 \\
4399.79 & 1.24 & $-$1.27 & D$-$ & MFW & & 139 & 4.86 & +0.25 \\
4407.68 & 1.22 & $-$2.47 & D$-$ & MFW & & 15 & 4.82 & +0.16 \\%
\hline
\end{tabular}
\end{table*}

\begin{table*}
\noindent {\bf Table A1} (cont.) \\[1mm]
\setlength{\tabcolsep}{.0625cm}
\begin{tabular}{lrrlrlrrr}
\hline\hline
\rule[-2mm]{0mm}{1mm}{}\\[-4.mm]
$\lambda\,$(\AA) & $\chi\,(\mbox{eV})$ & $\log gf$ & Acc. & Src. & & 
$W_{\lambda}$(m\AA) & $\log \varepsilon$ &
$\Delta \log \varepsilon$ \\[.8mm]
\hline
Ti\,{\sc ii}:\\
4421.90 & 2.05 & $-$1.39 & X & KB & & 27 & 4.73 & +0.22 \\%
4443.78 & 1.08 & $-$0.70 & D$-$ & MFW & & 237 & 4.62 & +0.23 \\
4450.50 & 1.08 & $-$1.45 & D$-$ & MFW & & 98 & 4.75 & +0.23 \\
4468.52 & 1.13 & $-$0.60 & D$-$ & MFW & & 251 & 4.60 & +0.29 \\
4501.27 & 1.11 & $-$0.75 & D$-$ & MFW & & 217 & 4.61 & +0.26 \\
4533.97 & 1.24 & $-$0.77 & D$-$ & MFW & & 327 & 4.81 & +0.10 \\%
4563.77 & 1.22 & $-$0.96 & D$-$ & MFW & & 224 & 4.83 & +0.19 \\
4568.31 & 1.22 & $-$2.65 & D & MFW & & 9 & 4.83 & +0.19 \\%
4571.96 & 1.57 & $-$0.53 & D$-$ & MFW & & 264 & 4.81 & +0.15 \\
4708.67 & 1.24 & $-$2.21 & D & MFW & & 18 & 4.78 & +0.27 \\%
4763.81 & 1.22 & $-$2.45 & X & KB & & S & 4.86 & +0.15 \\%
4798.53 & 1.08 & $-$2.43 & X & KB & & S & 4.78 & +0.27 \\
4874.01 & 3.09 & $-$0.79 & D & MFW & & 41 & 4.83 & +0.17 \\%
4911.19 & 3.12 & $-$0.34 & D & MFW & & 56 & 4.66 & +0.18 \\
5069.09 & 3.12 & $-$1.39 & D & MFW & & 10 & 4.79 & +0.18 \\%
5072.28 & 3.12 & $-$0.75 & D & MFW & & 32 & 4.73 & +0.17 \\%
5129.15 & 1.89 & $-$1.39 & D & MFW & & 47 & 4.89 & +0.18 \\
5185.91 & 1.89 & $-$1.61 & D & MFW & & 81 & 4.84 & +0.23 \\%
5188.68 & 1.58 & $-$1.21 & D & MFW & & 100 & 4.83 & +0.27 \\
5336.77 & 1.58 & $-$1.70 & D & MFW & & S & 4.88 & +0.22 \\
5381.02 & 1.57 & $-$2.08 & D & MFW & & 20 & 4.88 & +0.17 \\[1.5mm]%
V\,{\sc ii}:\\
3916.42 & 1.43 & $-$1.06 & B & MFW & & S & 3.60 & {\ldots} \\
3951.97 & 1.48 & $-$0.74 & B & MFW & & 67 & 3.53 & {\ldots} \\
4005.71 & 1.82 & $-$0.46 & D & MFW & & S & 3.59 & {\ldots} \\
4023.39 & 1.80 & $-$0.52 & X & KB & & 60 & 3.57 & {\ldots} \\
4035.63 & 1.79 & $-$0.62 & X & KB & & 53 & 3.58 & {\ldots} \\
4036.78 & 1.47 & $-$1.54 & D & MFW & & 15 & 3.61 & {\ldots} \\%
4065.07 & 3.78 & $-$0.24 & X & KB & & 10 & 3.62 & {\ldots} \\%
4183.44 & 2.05 & $-$0.95 & X & KB & & S & 3.57 & {\ldots} \\[1.5mm]
Cr\,{\sc i}:\\
4274.80 & 0.00 & $-$0.23 & B & MFW & & S & 5.67 & {\ldots} \\%
4539.76 & 2.54 & $-$1.15 & B & MFW & & 25 & 5.81 & {\ldots} \\%
5204.51 & 0.94 & $-$0.20 & B & MFW & & S & 5.65 & {\ldots} \\%
5206.04 & 0.94 & 0.02 & B & MFW & & S & 5.58 & {\ldots} \\%
5208.41 & 0.94 & 0.16 & B & MFW & & 12 & 5.71 & {\ldots} \\[1.5mm]%
Cr\,{\sc ii}:\\
3979.51 & 5.65 & $-$0.73 & X & KB & & S(73) & 5.59 & {\ldots} \\
4037.97 & 6.49 & $-$0.56 & X & KB & & 41 & 5.60 & {\ldots} \\
4054.08 & 3.11 & $-$2.47 & X & KB & & S & 5.64 & {\ldots} \\%
4072.56 & 3.70 & $-$2.41 & X & KB & & 28 & 5.60 & {\ldots} \\%
4086.13 & 3.71 & $-$2.42 & X & KB & & 24 & 5.51 & {\ldots} \\
4132.11 & 11.47 & $-$2.19 & X & KB & & S & 5.65 & {\ldots} \\%
4132.42 & 3.74 & $-$2.35 & X & KB \\%
4145.78 & 5.30 & $-$1.16 & X & KB & & 77 & 5.77 & {\ldots} \\%
4207.34 & 3.81 & $-$2.48 & X & KB & & 19 & 5.48 & {\ldots} \\%
4242.36 & 3.87 & $-$1.33 & X & KB & & 192 & 5.71 & {\ldots} \\
4252.63 & 3.84 & $-$2.02 & X & KB & & S & 5.72 & {\ldots} \\
4261.91 & 3.87 & $-$1.53 & X & KB & & 148 & 5.66 & {\ldots} \\
4269.28 & 3.85 & $-$2.17 & X & KB & & 52 & 5.67 & {\ldots} \\
4275.57 & 3.86 & $-$1.71 & X & KB & & 104 & 5.71 & {\ldots} \\
4284.19 & 3.86 & $-$1.86 & X & KB & & 95 & 5.72 & {\ldots} \\
4362.92 & 5.64 & $-$1.89 & X & KB & & S & 5.77 & {\ldots} \\%
4555.01 & 4.07 & $-$1.38 & D & MFW & & S & 5.68 & {\ldots} \\
4565.74 & 4.04 & $-$2.11 & D & MFW & & 53 & 5.79 & {\ldots} \\%
4588.22 & 4.07 & $-$0.63 & D & MFW & & 286 & 5.62 & {\ldots} \\
\hline
\end{tabular}
\end{table*}

\begin{table*}
\noindent {\bf Table A1} (cont.) \\[1mm]
\setlength{\tabcolsep}{.0625cm}
\begin{tabular}{lrrlrlrrr}
\hline\hline
\rule[-2mm]{0mm}{1mm}{}\\[-4.mm]
$\lambda\,$(\AA) & $\chi\,(\mbox{eV})$ & $\log gf$ & Acc. & Src. & & 
$W_{\lambda}$(m\AA) & $\log \varepsilon$ &
$\Delta \log \varepsilon$ \\[.8mm]
\hline
Cr\,{\sc ii}:\\
4592.07 & 4.07 & $-$1.22 & D & MFW & & 141 & 5.52 & {\ldots} \\
4616.64 & 4.07 & $-$1.29 & D & MFW & & S & 5.50 & {\ldots} \\
4618.82 & 4.07 & $-$1.11 & D & MFW & & 222 & 5.75 & {\ldots} \\
4634.10 & 4.07 & $-$1.24 & D & MFW & & 184 & 5.76 & {\ldots} \\
4812.34 & 3.86 & $-$1.99 & X & KB & & 61 & 5.61 & {\ldots} \\
4824.12 & 3.87 & $-$0.96 & X & KB & & 244 & 5.62 & {\ldots} \\
4836.22 & 3.86 & $-$1.99 & X & KB & & 68 & 5.62 & {\ldots} \\
4836.94 & 3.09 & $-$1.13 & B & MFW \\%
4848.24 & 3.86 & $-$1.14 & X & KB & & S(224) & 5.59 & {\ldots} \\
4876.41 & 3.86 & $-$1.46 & D & MFW & & 178 & 5.61 & {\ldots} \\
4884.58 & 3.86 & $-$2.08 & D & MFW & & 53 & 5.60 & {\ldots} \\
4901.62 & 6.49 & $-$0.83 & X & KB & & 25 & 5.57 & {\ldots} \\
5237.33 & 4.06 & $-$1.16 & D & MFW & & 195 & 5.54 & {\ldots} \\
5246.76 & 3.71 & $-$2.45 & D & MFW & & 23 & 5.61 & {\ldots} \\
5310.69 & 4.07 & $-$2.28 & D & MFW & & 25 & 5.63 & {\ldots} \\%
5313.61 & 4.07 & $-$1.65 & D & MFW & & 101 & 5.62 & {\ldots} \\
5334.87 & 4.05 & $-$1.56 & X & KB & & 81 & 5.48 & {\ldots} \\
5420.91 & 3.76 & $-$2.36 & D & MFW & & 26 & 5.50 & {\ldots} \\
5478.37 & 4.16 & $-$1.91 & X & KB & & 63 & 5.68 & {\ldots} \\%
5508.63 & 4.16 & $-$2.11 & D & MFW & & 39 & 5.62 & {\ldots} \\
5620.63 & 6.46 & $-$1.14 & X & KB & & 12 & 5.62 & {\ldots} \\%
6053.47 & 4.74 & $-$2.16 & D & MFW & & 16 & 5.60 & {\ldots} \\[1.5mm]%
Mn\,{\sc i}:\\
4034.48 & 0.00 & $-$0.81 & C+ & MFW & & S & 5.40 & {\ldots} \\[1.5mm]%
Mn\,{\sc ii}:\\
4206.37 & 5.40 & $-$1.57 & X & KB & & 17 & 5.34 & {\ldots} \\
4326.16 & 5.40 & $-$1.25 & X & KB & & 30 & 5.42 & {\ldots} \\
4365.22 & 6.57 & $-$1.35 & X & KB & & S & 5.39 & {\ldots} \\%
4478.64 & 6.65 & $-$0.95 & X & KB & & 12 & 5.42 & {\ldots} \\
4755.73 & 5.40 & $-$1.24 & X & KB & & 29 & 5.40 & {\ldots} \\
4784.63 & 6.57 & $-$1.51 & X & KB & & 5 & 5.41 & {\ldots} \\%
4806.82 & 5.42 & $-$1.56 & X & KB & & 10 & 5.28 & {\ldots} \\%
5297.00 & 9.86 & $-$0.21 & X & KB & & 7 & 5.40 & {\ldots} \\
5297.03 & 9.86 & 0.43 & X & KB \\
5297.06 & 9.86 & 0.62 & X & KB \\
5302.40 & 9.87 & 0.23 & X & KB & & S & 5.43 & {\ldots} \\%
5302.43 & 9.87 & 1.00 & X & KB \\%
5559.05 & 6.19 & $-$1.32 & X & KB & & S & 5.41 & {\ldots} \\%
5570.54 & 6.18 & $-$1.44 & X & KB & & S & 5.42 & {\ldots} \\[1.5mm]%
Fe\,{\sc i}:\\
3872.50 & 0.99 & $-$0.92 & B+ & FMW & & 139 & 7.11 & {\ldots} \\%
3872.92 & 2.72 & $-$1.75 & C & FMW \\%
3895.66 & 0.11 & $-$1.67 & B+ & FMW & & S & 7.16 & {\ldots} \\%
3899.71 & 0.09 & $-$1.53 & B+ & FMW & & S & 7.21 & {\ldots} \\%
3918.42 & 2.79 & $-$1.01 & X & KB & & 57 & 7.16 & {\ldots} \\%
3922.91 & 0.05 & $-$1.65 & B+ & FMW & & 30 & 7.16 & {\ldots} \\
3927.92 & 0.11 & $-$1.59 & C & FMW & & 37 & 7.23 & {\ldots} \\
3930.30 & 0.09 & $-$1.59 & C & FMW & & S & 7.23 & {\ldots} \\%
4045.81 & 1.48 & 0.28 & B+ & FMW & & 132 & 7.16 & {\ldots} \\
4063.59 & 1.56 & 0.07 & C+ & FMW & & 94 & 7.13 & {\ldots} \\
4071.74 & 1.61 & -0.65 & B+ & FMW & & 70 & 7.15 & {\ldots} \\
4147.34 & 3.32 & $-$1.90 & X & KB & & 11 & 7.21 & {\ldots} \\%
4147.49 & 3.32 & $-$2.47 & B+ & FMW \\%
4147.67 & 1.48 & $-$2.10 & B+ & FMW \\%
4154.50 & 2.79 & $-$0.48 & X & KB & & S & 7.21 & {\ldots} \\%
4154.81 & 3.33 & $-$0.37 & C+ & FMW \\%
\hline
\end{tabular}
\end{table*}

\begin{table*}
\noindent {\bf Table A1} (cont.) \\[1mm]
\setlength{\tabcolsep}{.0625cm}
\begin{tabular}{lrrlrlrrr}
\hline\hline
\rule[-2mm]{0mm}{1mm}{}\\[-4.mm]
$\lambda\,$(\AA) & $\chi\,(\mbox{eV})$ & $\log gf$ & Acc. & Src. & & 
$W_{\lambda}$(m\AA) & $\log \varepsilon$ &
$\Delta \log \varepsilon$ \\[.8mm]
\hline
Fe\,{\sc i}:\\
4181.75 & 2.83 & $-$0.18 & D$-$ & FMW & & 15 & 7.16 & {\ldots} \\
4219.36 & 3.57 & 0.12 & C+ & FMW & & 8 & 7.21 & {\ldots} \\%
4278.23 & 3.37 & $-$1.74 & C & FMW & & 75 & 7.31 & {\ldots} \\
4404.75 & 1.56 & $-$0.14 & B$+$ & FMW & & 63 & 7.19 & {\ldots} \\
4466.55 & 2.83 & $-$0.59 & C+ & FMW & & S & 7.31 & {\ldots} \\%
4476.02 & 2.79 & $-$0.73 & X & KB & & S & 7.26 & {\ldots} \\%
4476.08 & 3.59 & $-$0.37 & X & KB \\%
4871.32 & 2.87 & $-$0.41 & C+ & FMW & & S(39) & 7.21 & {\ldots} \\%
4918.95 & 4.10 & $-$0.67 & X & KB & & 11 & 7.23 & {\ldots} \\%
4918.99 & 2.87 & $-$0.37 & C+ & FMW \\[1.5mm]%
Fe\,{\sc ii}:\\
3938.29 & 1.67 & $-$3.89 & D & FMW & & S & 7.23 & $-$0.03 \\
4031.44 & 4.73 & $-$3.12 & X & KB & & S & 7.21 & +0.02 \\
4051.21 & 5.52 & $-$2.99 & X & KB & & S & 7.26 & $\pm$0.00 \\%
4122.64 & 2.58 & $-$3.38 & D & FMW & & 176 & 7.36 & $\pm$0.00 \\
4173.46 & 2.58 & $-$2.18 & C & FMW & & 328 & 7.16 & +0.05 \\%
4273.32 & 2.70 & $-$3.34 & D & FMW & & 152 & 7.26 & $\pm$0.00 \\
4296.57 & 2.70 & $-$3.01 & D & FMW & & 228 & 7.21 & $-$0.05 \\
4303.17 & 2.70 & $-$2.49 & C & FMW & & 284 & 7.25 & $-$0.01 \\
4385.39 & 2.78 & $-$2.57 & D & FMW & & S & 7.25 & $\pm$0.00 \\%
4472.62 & 7.65 & $-$2.34 & X & KB & & 102 & 7.24 & +0.01 \\%
4489.19 & 2.83 & $-$2.23 & D & FMW & & 186 & 7.31 & $\pm$0.00 \\
4491.40 & 2.86 & $-$2.70 & C & FMW & & 233 & 7.26 & $-$0.02 \\
4508.28 & 2.86 & $-$2.31 & D & KB & & 346 & 7.24 & $-$0.02 \\
4515.34 & 2.84 & $-$2.48 & D & FMW & & 276 & 7.26 & $-$0.03 \\
4520.23 & 2.81 & $-$2.60 & D & FMW & & 314 & 7.31 & $-$0.05 \\
4541.52 & 2.86 & $-$3.05 & D & FMW & & 172 & 7.31 & $-$0.05 \\
4576.33 & 2.84 & $-$3.04 & D & FMW & & 212 & 7.31 & $-$0.05 \\
4620.51 & 2.83 & $-$3.28 & D & FMW & & 142 & 7.23 & $-$0.08 \\
4656.97 & 2.89 & $-$3.63 & E & FMW & & 73 & 7.23 & $-$0.08 \\
4993.35 & 2.81 & $-$3.65 & E & FMW & & S & 7.21 & $-$0.05 \\
5074.05 & 6.75 & $-$1.97 & X & KB & & 30 & 7.26 & $-$0.05 \\%
5197.48 & 5.89 & $-$2.72 & X & KB & & 317 & 7.21 & $-$0.05 \\%
5197.58 & 3.23 & $-$2.79 & C & FMW \\%
5254.93 & 3.14 & $-$3.23 & X & KB & & 109 & 7.26 & $-$0.05 \\%
5276.00 & 3.20 & $-$1.94 & C & FMW & & 385 & 7.21 & $-$0.10 \\%
5427.83 & 6.72 & $-$1.66 & X & KB & & 25 & 7.21 & $-$0.02 \\
6147.74 & 3.89 & $-$2.72 & X & KB & & 140 & 7.39 & $-$0.02 \\%
6149.26 & 3.89 & $-$2.72 & X & KB & & 126 & 7.39 & $-$0.02 \\
6238.39 & 3.89 & $-$2.63 & X & KB & & 135 & 7.31 & $-$0.08 \\[1.5mm]
Fe\,{\sc iii}:\\
4419.60 & 8.24 & $-$2.22 & X & KB & & 10 & 7.28 & {\ldots} \\[1.5mm]
Ni\,{\sc ii}:\\
4015.47 & 4.03 & $-$2.42 & X & KB & & 76 & 6.19 & {\ldots} \\
4067.03 & 4.03 & $-$1.84 & X & KB & & 163 & 6.12 & {\ldots} \\
4192.07 & 4.03 & $-$3.06 & X & KB & & 26 & 6.22 & {\ldots} \\
4244.78 & 4.03 & $-$3.11 & X & KB & & 22 & 6.19 & {\ldots} \\
4362.10 & 4.03 & $-$2.72 & X & KB & & 44 & 6.21 & {\ldots} \\
4679.16 & 6.84 & $-$1.75 & X & KB & & 14 & 6.26 & {\ldots} \\%
5065.98 & 12.29 & 0.01 & X & KB & & 10 & 6.13 & {\ldots} \\%
5066.06 & 14.33 & $-$0.83 & X & KB \\[1.5mm]%
Sr\,{\sc ii}:\\
4077.71 & 0.00 & 0.15 & X & FW & & 104 & 2.00 & {\ldots} \\
4215.52 & 0.00 & $-$0.17 & X & FW & & 77 & 2.05 & {\ldots} \\[1.5mm]
Ba\,{\sc ii}:\\
4554.03 & 0.00 & 0.14 & X & D & & S & 2.03 & {\ldots} \\
\hline
\end{tabular}
\end{table*}

\begin{table*}
\begin{minipage}{0.45\textwidth}
accuracy indicators $-$ uncertainties within: AA: 1\%; A: 3\%; B: 10\%; C:
25\%; D: 50\%; E: larger than 50\%; X: unknown\\[1mm] 
sources of $gf$-values $-$ BMZ: Butler et al.~(\cite{butler1993}); CA:
Coulomb approximation (Bates \& Damgaard~\cite{bates1949});
D: Davidson et al.~(\cite{davidson1992});
F:~Fernley et al.~(available from {\sc Topbase});
FMW: Fuhr et al.~(\cite{fuhr1988});
KB: Kurucz \& Bell~(\cite{kurucz1995});
MEL: Mendoza et al.~(available from {\sc Topbase});
MFW:~Martin et al.~(\cite{martin1988});
S: Sigut~(\cite{sigut1999});
T: Taylor~(available from {\sc Topbase});
WFD: Wiese et al.~(\cite{wiese1996});
WSG: Wiese et al.~(\cite{wiese1966});
WSM:~Wiese et al.~(\cite{wiese1969});
when available$^{(*)}$, improved $gf$-values from Fuhr \& Wiese~(\cite{fuhr1998}) 
are favoured\\[1mm]
sources for Stark broadening parameters $-$ 
\ion{H}{i}: Stehl\'e \& Hutcheon~(\cite{stehle1999}), Vidal et al.~(\cite{vidal1973});
\ion{He}{i}: Barnard et al.~(\cite{barnard1969}), 
Dimitrijevi\'c \& Sahal-Br\'echot~(\cite{dimitrijevic1990}); 
\ion{C}{i}: Griem~(\cite{griem1974}),~Cowley (\cite{cowley1971});
\ion{C}{ii}: Griem~(\cite{griem1964}, \cite{griem1974}), Cowley~(\cite{cowley1971});
\ion{N}{i/ii}: Griem~(\cite{griem1964}, \cite{griem1974}), Cowley~(\cite{cowley1971});
\ion{O}{i}: Cowley~(\cite{cowley1971});
\ion{Ne}{i}: Griem~(\cite{griem1974}), Cowley~(\cite{cowley1971});
\ion{Na}{i}: Cowley~(\cite{cowley1971}),~Griem~(\cite{griem1974});
\ion{Mg}{i}: Dimitrijevi\'c \& Sahal-Br\'echot~(\cite{dimitrijevic1996}), Cowley~(\cite{cowley1971});
\ion{Mg}{ii}: Griem~(\cite{griem1964}, \cite{griem1974}), Cowley~(\cite{cowley1971});
\ion{Al}{i}: Griem~(\cite{griem1974}), Cowley~(\cite{cowley1971});
\ion{Al}{ii}: Griem~(\cite{griem1964}, \cite{griem1974}), Cowley~(\cite{cowley1971});
\ion{Si}{ii}: Lanz et al.~(\cite{lanz1988}), Griem~(\cite{griem1974}), Cowley~(\cite{cowley1971});
\ion{S}{ii}: Cowley~(\cite{cowley1971});
\ion{Ca}{ii}: Griem~(\cite{griem1974}), Cowley~(\cite{cowley1971});
Sc $-$ Ni: Cowley~(\cite{cowley1971});
\ion{Sr}{ii}: Cowley~(\cite{cowley1971});
\ion{Ba}{ii}: Dimitrijevi\'c \& Sahal-Br\'echot~(\cite{dimitrijevic1997})
\end{minipage}
\end{table*}

\end{multicols}{2}

\end{document}